\documentclass[aps,pre,twocolumn,longbibliography,superscriptaddress,nopacs,amsmath,amssymb,letterpaper]{revtex4-1}

\usepackage{graphicx}
\usepackage{amsmath, amsfonts, amsthm}
\usepackage{amssymb}
\usepackage[colorlinks=true, linkcolor=teal, urlcolor=teal, citecolor=teal]{hyperref}
\usepackage{mathtools}
\usepackage{mathrsfs}
\usepackage{bm}
\usepackage{xcolor}
\usepackage[caption=false]{subfig}
\usepackage{braket}

\usepackage[normalem]{ulem}

\begin{document}

\title{The emergence of informative higher scales in complex networks}

\author{Brennan Klein}
\email{klein.br@northeastern.edu}
\affiliation{Network Science Institute, Northeastern University, Boston, MA 02115, USA}
\affiliation{Laboratory for the Modeling of Biological and Socio-Technical Systems, Northeastern University, Boston, MA 02115, USA}

\author{Erik Hoel}
\email{erik.hoel@tufts.edu}
\affiliation{Allen Discovery Center, Tufts University, Medford, MA 02155, USA}

\date{\today}

\begin{abstract}
The connectivity of a network contains information about the relationships between nodes, which can denote interactions, associations, or dependencies. We show that this information can be analyzed by measuring the uncertainty (and certainty) contained in paths along nodes and links in a network. Specifically, we derive from first principles a measure known as \textit{effective information} and describe its behavior in common network models. Networks with higher effective information contain more information in the relationships between nodes. We show how subgraphs of nodes can be grouped into macro-nodes, reducing the size of a network while increasing its effective information (a phenomenon known as \textit{causal emergence}). We find that informative higher scales are common in simulated and real networks across biological, social, informational, and technological domains. These results show that the emergence of higher scales in networks can be directly assessed and that these higher scales offer a way to create certainty out of uncertainty.
\end{abstract}

\maketitle

\section{\label{sec:introduction}Introduction}

Networks provide a powerful syntax for representing a wide range of systems, from the trivially simple to the highly complex \cite{Barabasi2016, Newman2010, Amaral2004}. It is common to characterize networks based on structural properties like their degree distribution or clustering, and the study of such properties has been crucial for the growth of Network Science. Yet there remains a gap in our treatment of the information contained in the relationships between nodes in a network, particularly in networks that have both weighted connections and feedback, which are hallmarks of complex systems \cite{Koseska2017CellProcess, Rodrigues2016TheNetworks}. As we will show, analyzing this information allows for modeling the network at the most appropriate, informative scale. This is especially critical for networks that describe interactions or dependencies between nodes such as contact networks in epidemiology \cite{Perra2012ActivityNetworks}, neuronal and functional networks in the brain \cite{Bassett2017}, or interaction networks among cells, genes, or drugs \cite{Barabasi2011}, as these networks can often be analyzed at multiple different scales.

Here we introduce information-theoretic measures that capture the information contained in the connectivity of a network, which can be used to identify when these networks possess informative higher scales. To do so, we focus on the out-weight vector, $W^{out}_{i}$, of each node, $v_i$, in a network. This vector consists of weights $w_{ij}$ between $v_i$ and its neighbors, $v_j$, and $w_{ij} = 0$ if there is no edge from $v_i$ to $v_j$. For each $W^{out}_{i}$ we assume $\sum_j w_{ij} = 1$, which means $w_{ij}$ can be interpreted as the probability $p_{ij}$ that a random walker on $v_i$ will transition to $v_j$ in the next time step, where a random walker might represent the passing of a signal, an interaction, or a state-transition \cite{Masuda2017}. The information contained in a network's connectivity can be characterized by the uncertainty among its nodes' out-weights and in-weights. The total information in the relationships between nodes is a function of this uncertainty and can be derived from two properties. 

The first is the uncertainty of a node's outputs, which is the Shannon entropy \cite{Shannon1948} of its out-weights, $H(W^{out}_{i})$. The average of this entropy, $\langle H(W^{out}_{i} )\rangle $, across all nodes is the amount of noise present in the network's relationships. Only if $\langle H(W^{out}_{i} )\rangle =0$ is the network is \textit{deterministic}.

The second property is how weight is distributed across the whole network, $\langle W^{out}_{i}\rangle$. This vector is composed of elements that are the sum of the in-weights $w_{ji}$ to each node $v_i$ from each of its incoming neighbors, $v_j$ (then normalized by total weight of the network). Its entropy, $H (\langle W^{out}_{i}\rangle)$, reflects how certainty is distributed across the network. If all nodes link only to the same node, then $H(\langle W^{out}_{i}\rangle) = 0$, and the network is totally \textit{degenerate} since all nodes lead to the same node.

The \textit{effective information} ($EI$) of a network is the difference between these two quantities:
\begin{equation}\label{eq:ei}
    EI = H(\langle W^{out}_{i} \rangle) - \langle {H}(W^{out}_{i}) \rangle \tag{1}
\end{equation}

The entropy of the distribution of out-weights in the network forms an upper bound of the amount of unique information in the network's relationships, from which the information lost due to the uncertainty of those relationships is subtracted. Networks with high $EI$ contain more certainty in the relationships between nodes in the network (since the links represent less uncertain dependencies, unique associations, or deterministic transitions), whereas networks with low $EI$ contain less certainty. Note that $EI$ can be interpreted simply as a structural property of random walkers on a network and their behavior, similar to other common network measures \cite{Masuda2017}.

\begin{figure*}[t!]
    \centering
    \subfloat[\label{fig:ei_p}]{
        \includegraphics[width=1.0\columnwidth]{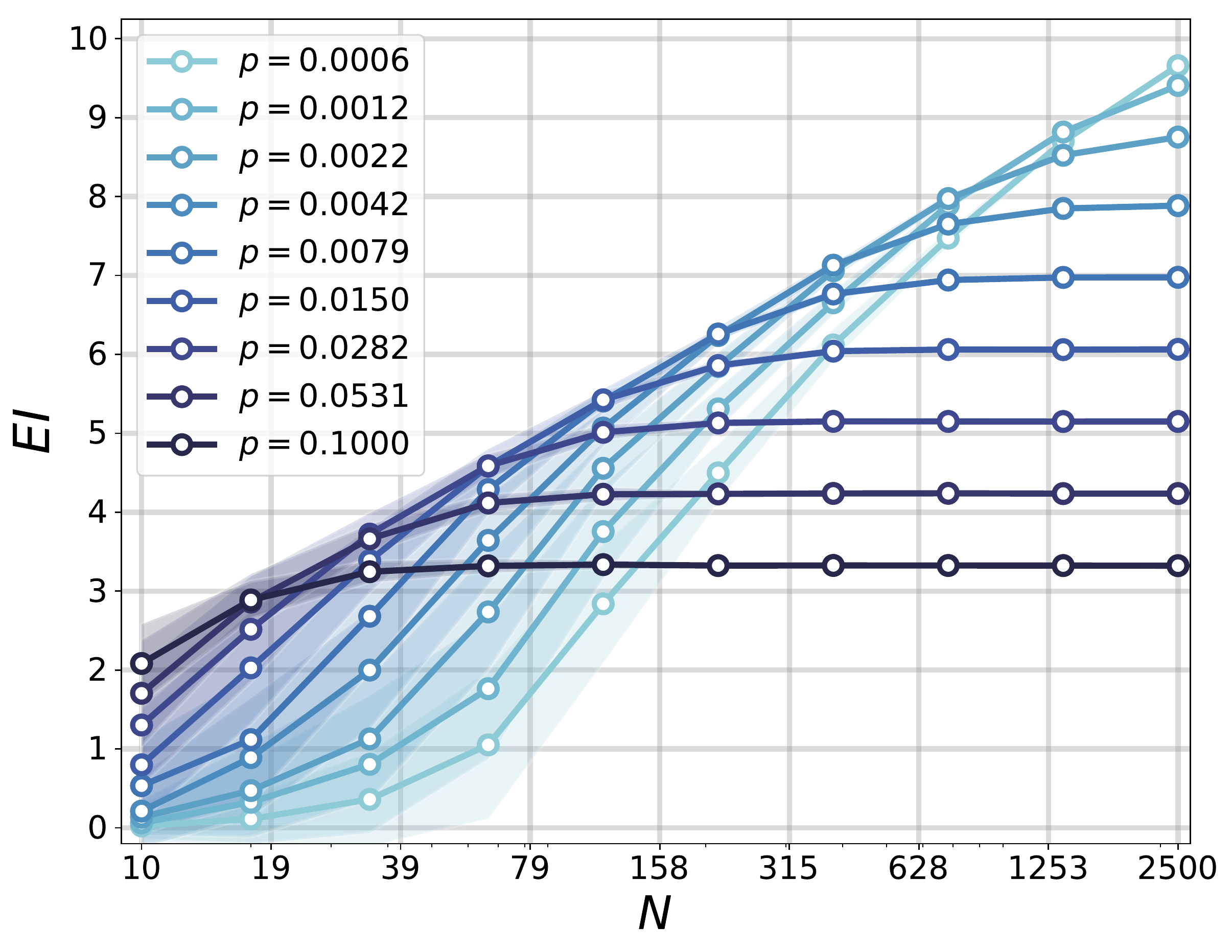}
    }\hfill
    \subfloat[\label{fig:ei_alpha}]{
        \includegraphics[width=1.0\columnwidth]{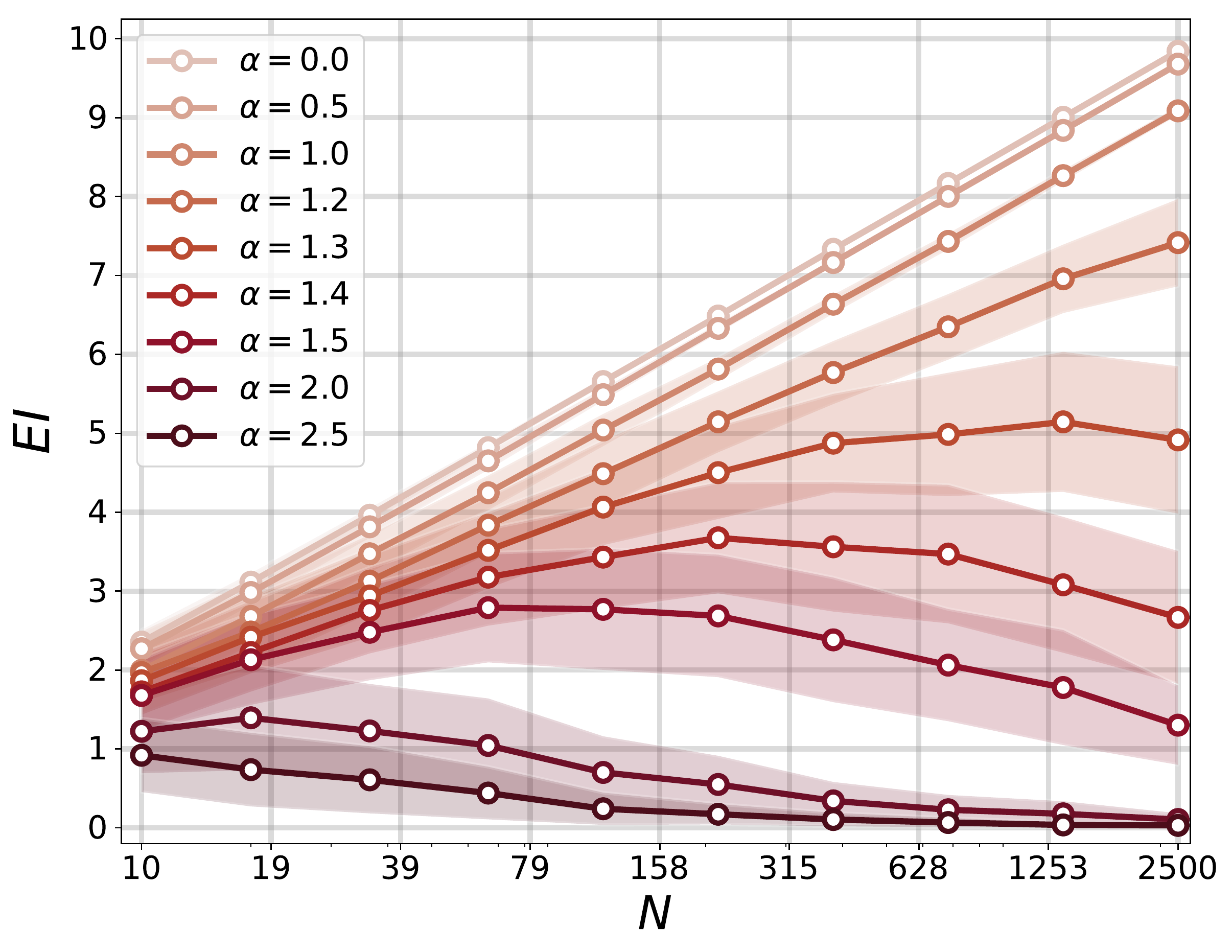}
    }
    \caption{\textbf{Effective information depends on network structure.} \textbf{(A)} In Erd\H{o}s-R\'enyi (ER) networks we see the network's $EI$ level off at $EI=-\log_2(p)$ as $N$, the network's size, increases (log scale shown). \textbf{(B)} The $EI$ of networks grown under a preferential attachment mechanism, which depends on the preferential attachment exponent, $\alpha$. Under this network growth model, new nodes add their $m$ edges (here, $m=1$) to existing nodes in the network with a probability proportional to $k^\alpha$. Only sublinear preferential attachment ($\alpha < 1.0$) allows for the continuous growth of $EI$ with the growth of the network. The ribbons around the data represent standard deviations after 100 simulations of each.}
    \label{fig:EI_results}
\end{figure*}

Here, we use this measure to develop a general classification of networks (key terms can be found in Supplementary Materials, SM \ref{sec:SI_keyTerms}). Furthermore, we show how the connectivity and different growth rules of a network have a deep relationship to that network's $EI$. This also provides a principled means of quantifying the amount of information among the micro-, meso-, and macroscale dependencies in a network. We introduce a formalism for finding and assessing the most informative scale of a network: the scale that minimizes the uncertainty in the relationships between nodes. For some networks, a macroscale description of the network can be more informative in this manner, demonstrating a phenomenon known as \textit{causal emergence} \cite{Hoel2013QuantifyingMicro, Hoel2018}, which here we generalize to complex networks. This provides a rigorous means of identifying when networks possess an informative higher scale.

\section{Results}
\subsection{\label{sec:intro_ei}Effective information quantifies a network's dependencies}
This work expands to networks previous research on using effective information to measure the amount of information in the causal relationships between the mechanisms or states of a system. Originally, $EI$ was introduced to capture the causal influence between two subsets of neurons as a step in the calculation of integrated information in the brain \cite{Tononi2003}. Later, a system-wide version of $EI$ was shown to capture fundamental causal properties in Boolean networks of logic gates, particularly their determinism and degeneracy \cite{Hoel2013QuantifyingMicro}.

Our current derivation from first principles of an $EI$ for networks is equivalent to this system-wide definition (SM \ref{sec:ei_calc}), which was based originally on interventions upon system states. For example, if a system in a particular state, \textit{A}, always transitions to state \textit{B}, the causal relationship between \textit{A} and \textit{B} can be represented by a node-link diagram wherein the two nodes---\textit{A} and \textit{B}---are connected by a directed arrow, indicating that \textit{B} depends on \textit{A}. This might be a node pair in a ``causal diagram'' (often represented as a directed acyclic graph, or a \textit{DAG}) such as those used in \cite{Pearl2000, Pearl1995} to represent interventions and causal relationships. In such a case, the information in the causal relationship between \textit{A} and \textit{B} can be assessed by intervening to randomize \textit{A} ($do(A=H^{max})$) and measuring the effects on \textit{B}. The $EI$ would be the mutual information between \textit{A} and \textit{B} under such randomization: $I(do(A=H^{max}),B)$ \cite{Hoel2017WhenTerritory}.

To expand this framework to networks in general, we relax this intervention requirement by assuming that the elements in $W^{out}_{i}$ sum to $1$. In this case, an ``intervention'' can be interpreted as dropping a random walker on the network. For example, if the network represents a DAG or Markov chain, then dropping a random walker on a node $v_i$ would be equivalent to $do(v_i)$. The entropy of the transitions of the random walkers and how those transitions are distributed defines the $EI$ of a network. In this generalized formulation, only in networks where the nodes and edges actually represent dynamics, interactions, or couplings does $EI$ indicate information about causation. In the case where edges represent correlations, or where what nodes or edges represent is undefined, $EI$ is merely a structural property of the information contained in the behavior of hypothetical random walkers (however, this situation is no different from other analysis methods that rely on random walkers).

Here we describe how this generalized structural $EI$ property behaves in common network models, asking basic questions about the relationship between a network's $EI$ and its size, density, and structure. These inquiries allow for the exhaustive classification and quantification of the information contained in the connectivity of real networks. It is intuitive that the $EI$ of a network will increase as the network grows in size. In general, adding more nodes should increase the entropy, which should in turn increase the amount of information. However, in cases of randomness rather than structure, $EI$ should reflect this randomness. We found this is indeed the case.

\begin{figure*}[t]
    \centering
    \subfloat[\label{fig:Determ_Degen_hist}]{
        \includegraphics[width=1.0\columnwidth]{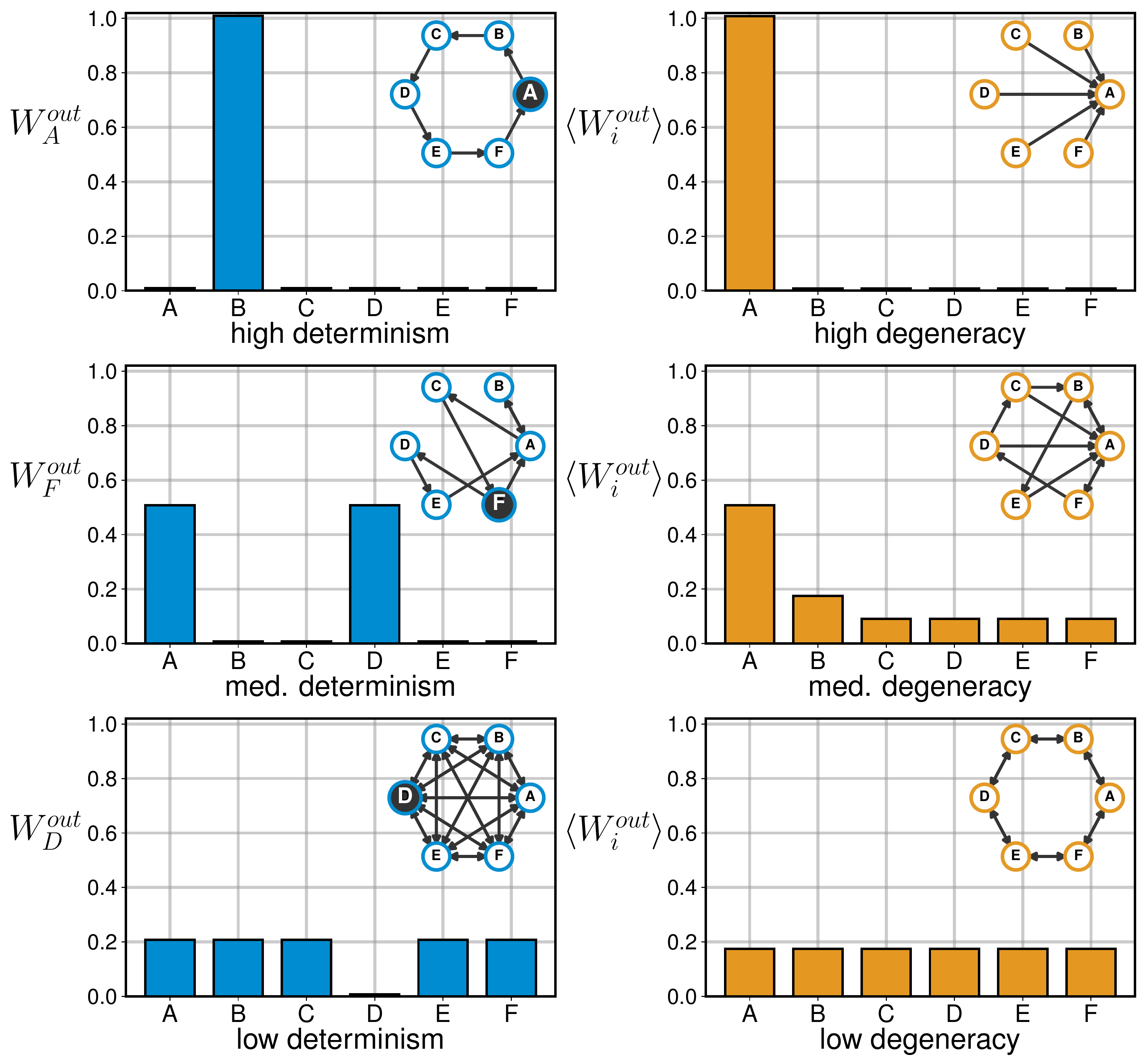}
    }\hfill
    \subfloat[\label{fig:Determ_Degen_Networks}]{
        \includegraphics[width=0.95\columnwidth]{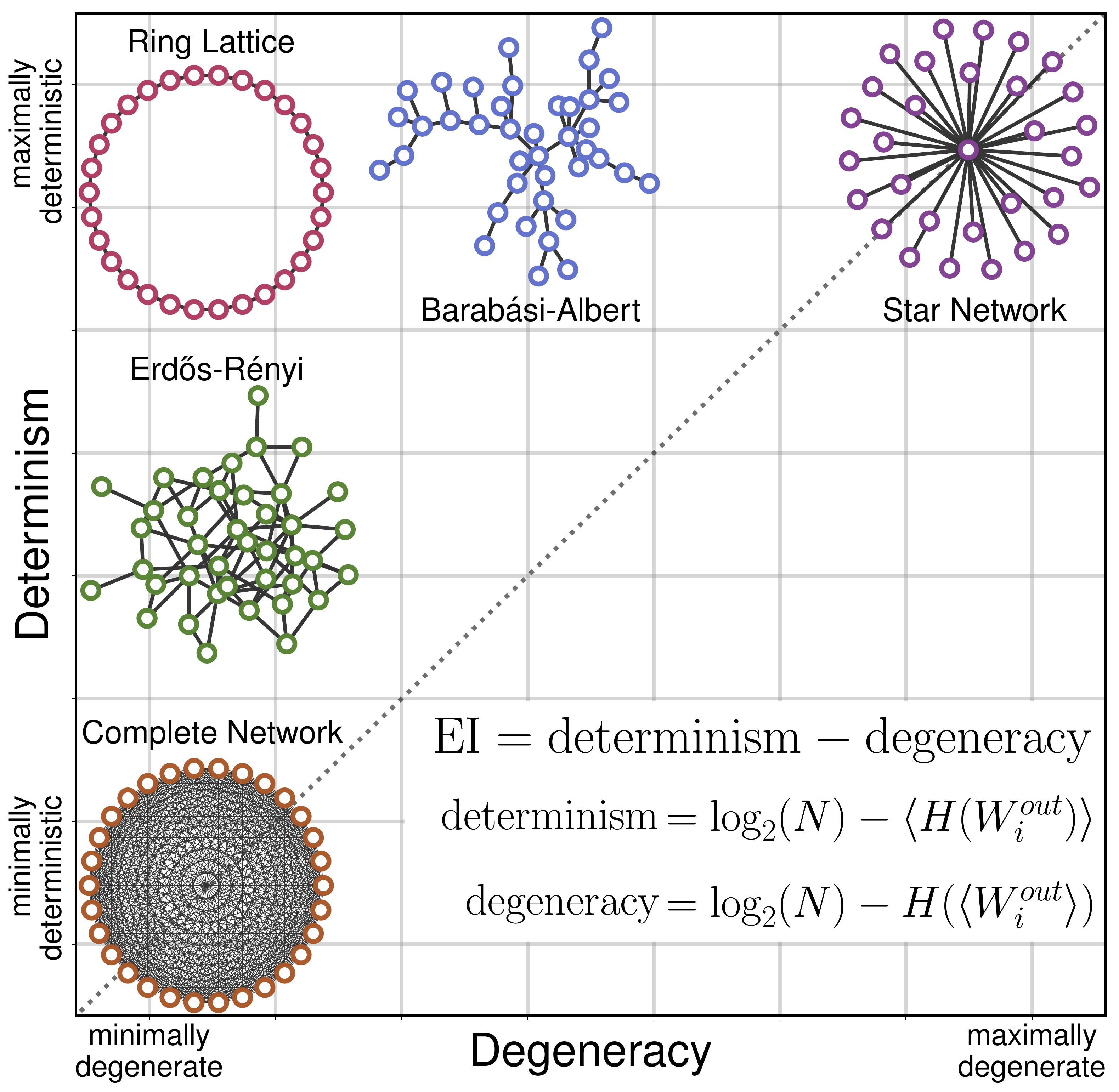}
    }
    \caption{\textbf{Comparing determinism and degeneracy.} \textbf{(A)} Left column: three example out-weight vectors, $W^{out}_{i}$, of a given node, $v_i$. A maximally deterministic vector (top left, where the $W^{out}_{A}$ corresponds to node $A$ in the inset network motif) is when a random walker on $v_i$ transitions to one of its neighbors with probability $1.0$, whereas indeterminism occurs when $v_i$ has a uniform probability of visiting any node in the network in the next time step. Right: three example in-weight vectors to a given $v_j$. A maximally degenerate vector, $\langle W^{out}_{i}\rangle$ (top right, exemplified by the inset network motif), is when every outgoing edge in the network connects to a single node, whereas minimal degeneracy occurs when each value in $\langle W^{out}_{i}\rangle$ is uniformly $\frac{1}{N}$. \textbf{(B)} By comparing the determinism and degeneracy of canonical network structures, we find a great deal of heterogeneity in different network models' ratios between their determinism and degeneracy. High degeneracy is characterized by hub-and-spoke topology, as in the case of the star network. Networks with high determinism are characterized by longer average path lengths, as in the case of a ring lattice.}
    \label{fig:determ_degen}
\end{figure*}

Fig. \ref{fig:ei_p} shows the relationship between a network's $EI$ and its size under several parameterizations of Erd\H{o}s-R\'enyi (ER) random graphs \cite{Erdos1959, Bollobas1984}. As the size of an ER network increases (while keeping constant the probability that any two nodes will be connected, $p$), its $EI$ converges to a value of $-\log_2(p)$. That is, in random networks $EI$ is dominated solely by the probability that any two nodes are connected, a key finding which demonstrates that, after a certain point, a random network structure does not contain more information as its size increases. This shift occurs in ER networks at approximately $\langle k \rangle = \log_2(N)$, which is also the point at which we can expect all nodes to be in a giant component \cite{Barabasi2016}. This finding illustrates that network connectivity must be non-random to increase the amount of information in the relationships between nodes (see SM \ref{sec:SI_derivations_ER} for derivation). Note that if a network is maximally dense (i.e. a fully connected network, with self-loops), $EI=0.0$. However, we expect such dense low-$EI$ structures to be uncommon, since network structures found in nature and society tend to be sparse \cite{DelGenio2011}.

We report another key relationship between a network's connectivity and its $EI$ in Fig. \ref{fig:ei_alpha}. We again compare the $EI$ of a network to its size, focusing on networks grown under different parameterizations of a preferential attachment model \cite{Krapivsky2000, Barabasi1999}. Under a preferential attachment growth model, a new node is added to the network at each time step, contributing $m$ new edges to the network; these $m$ edges connect to nodes already in the network, $v_j$, with a probability proportional to $k_j^\alpha$. Here $k_j$ is the degree of node $v_j$ and $\alpha$ tunes the amount of preferential attachment. A value of $\alpha=0.0$ corresponds to each node having an equal chance of receiving a new node's link (i.e., no preferential attachment). The classic Barabási-Albert network corresponds to linear preferential attachment, $\alpha=1.0$ \cite{Barabasi1999}. Superlinear preferential attachment, $\alpha > 1.0$, creates networks that have less and less $EI$, eventually resembling star-like structures (see SM \ref{sec:SI_derivations_ring} for derivation). As shown in Fig. \ref{fig:ei_alpha}, only in cases of sublinear preferential attachment, $\alpha < 1.0$, does the network's $EI$ continue to increase with its size. When $\alpha=0.0$---creating a random tree---the network's $EI$ increases logarithmically as its size increases.

The maximum possible $EI$ in a network of $N$ nodes is $\log_2(N)$. This can be seen in the case of a directed ring network where each node has one incoming link and one outgoing link, each with a weight of $1.0$, so each node has one node uniquely connecting to it. In such a network, each node contributes zero uncertainty, since ${\langle H(W^{out}_{i}) \rangle}=0.0$, and ${H} (\langle W^{out}_{i}\rangle) =\log_2(N)$, and therefore its $EI$ is always $\log_2(N)$. In general, the $EI$ of undirected lattices is fixed entirely by its size and the dimension of the ring lattice (i.e. $d=1$ is an undirected ring, $d=2$ is a taurus, etc. \cite{Watts1998}), so for such lattices $EI = \log_2(N) - \log_2(2d)$ (see SM \ref{sec:SI_derivations_ring} for derivation).

The picture that emerges is that $EI$ is inextricably linked a network's connectivity and growth (even network motifs, as shown in SM \ref{sec:motifs}) and therefore to the fundamentals of Network Science. Random networks have a fixed amount of $EI$, and scale-freeness ($\alpha=1.0$) represents the critical bound for the growth of $EI$. In general, dense networks and star-like networks have less $EI$. The next section explores how $EI$'s components explain these associations.

\subsection{\label{sec:determdegen}Determinism and degeneracy}

\textit{Determinism} and \textit{degeneracy} are the two fundamental components of $EI$ \cite{Hoel2013QuantifyingMicro}. They are based on a network's connectivity (see Figure  \ref{fig:Determ_Degen_hist} for a visual explanation), specifically the degree of overlapping weight in the networks. Determinism and degeneracy are derived from the uncertainty over outputs and uncertainty in how those outputs are distributed, respectively:
\begin{align} 
    \text{determinism} &= \log_2(N) - \langle H(W^{out}_{i}) \rangle \tag{2} \\
    \text{degeneracy} &= \log_2(N) - H(\langle W^{out}_{i}\rangle) \tag{3}
\end{align}

In a maximally deterministic network wherein all nodes have a single output, $w_{ij}=1.0$, the determinism is $\log_2(N)$ because $\langle H(W^{out}_{i})\rangle =0.0$. Conceptually, this means that a random walker will move deterministically starting from any node. Degeneracy is the amount of information in the connectivity lost via an overlap in input weights (e.g., if multiple nodes output to the same node). In a perfectly non-degenerate system where all nodes have equal input weights, the degeneracy is zero since ${H}(\langle W^{out}_{i}\rangle) = \log_2(N)$. Together, determinism and degeneracy can be used to define $EI$:
\begin{equation} \label{eq:effective_info_det_degen}
    EI = \text{determinism} - \text{degeneracy}
    \tag{4}
\end{equation}

These two quantities provide clear explanations for why different networks have the $EI$ they do. For example, as the size of an Erd\H{o}s-R\'enyi random network increases, its degeneracy approaches zero, which means the $EI$ of a random network is driven only by the determinism of the network, which is in turn the negative log of the probability of connection, \textit{p}. Similarly, in \textit{d}-dimensional ring lattice networks, the degeneracy term is always zero, which means the $EI$ of a ring lattice structure also reduces to the determinism of that structure. Ring networks with an average degree $\langle k \rangle$ will have a higher $EI$ than ER networks with the same average degree because ring networks will have a higher determinism value. In the case of star networks, the degeneracy term alone governs the decay of the $EI$ such that hub-and-spoke-like structures quickly become uninformative in terms of cause and effect (see SM \ref{sec:SI_derivations} for derivations concerning these cases).  In general, this means that canonical networks can be characterized by their ratio of determinism to degeneracy (see Fig. \ref{fig:Determ_Degen_Networks}).

\subsection{\label{sec:EI_real_networks}Effective information in real networks}

So far, we have been agnostic as to the origin of the network under analysis. As described previously, to measure the $EI$ of a network one can create each $W^{out}_{i}$ by normalizing each node's out-weight vector to sum to $1.0$. Regardless of what the relationships between the nodes represent, the network's determinism reflects how targeted the out-weights of the nodes are (networks with more targeted links possess higher $EI$), while the degeneracy captures the overlap of the targeting of nodes. High $EI$ reflects the greater specificity in the connectivity, whereas low $EI$ indicates a lack of specificity (as in Fig. \ref{fig:Determ_Degen_hist}). This generalizes our results to multiple types of representations, although the origin of the normalized network should be kept in mind when interpreting the value of the measure.

\begin{figure}[t!]
    \centering
    \includegraphics[width=1.0\columnwidth]{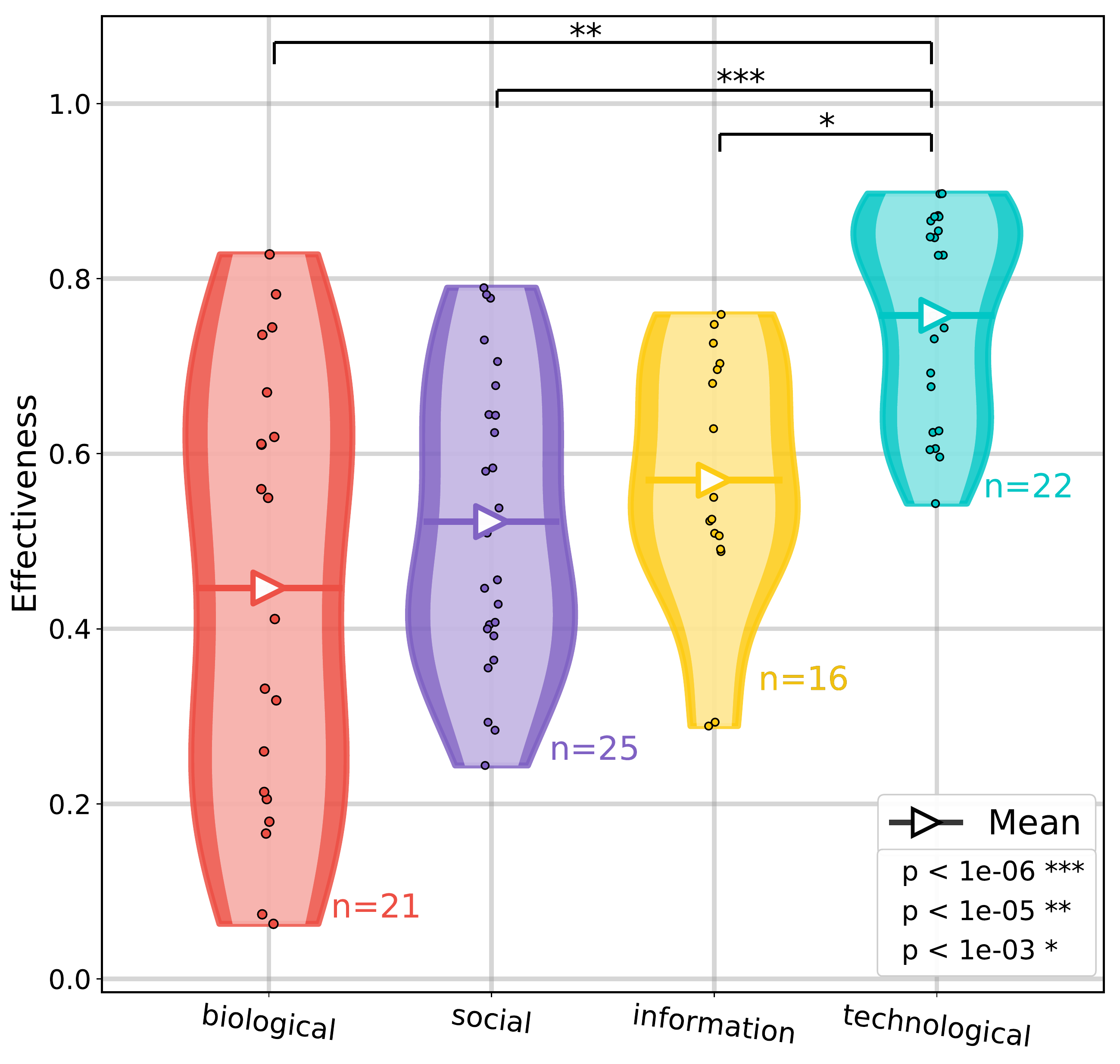}
    \caption{\textbf{Effective information of real networks.} Effectiveness, a network's $EI$, normalized by $\log_2(N)$ \cite{Hoel2013QuantifyingMicro}, of 84 real networks from the Konect network database \cite{Kunegis2013}, grouped by domain of origin. To look further at the names and domains of the networks in question, see SM \ref{sec:SI_real_network_table}. Networks in different categories have varying effectiveness (t-test, comparison of means).}
    \label{fig:ei_real_nets}
\end{figure}

Since the $EI$ of a network will change depending on the network's size, we use a normalized form of $EI$ known as \textit{effectiveness} in order to compare the $EI$ of real networks. Effectiveness ranges from $0.0$ to $1.0$ and is defined as:
\begin{equation}\label{eq:effectiveness}
    \text{effectiveness} = \frac{EI}{\log_2(N)}\tag{5}
\end{equation}

As the determinism and degeneracy of a network increase to their minimum and maximum possible values, respectively, the effectiveness of that network will trend to $0.0$. Regardless of its size, a network wherein each node has a deterministic output to a unique target has an effectiveness of $1.0$. 

In Fig. \ref{fig:ei_real_nets}, we examine the effectiveness of 84 different networks corresponding to data from real systems. These networks were selected primarily from the Konect Network Database \cite{Kunegis2013}, which was used because its networks are publicly available, range in size from dozens to tens of thousands of nodes, often have a reasonable interpretation as being based on interactions between nodes, and they are diverse, ranging from social networks, to power networks, to metabolic networks. We defined four categories of interest: biological, social, informational, and technological. We selected our networks by using all the available networks (under 40,000 nodes due to computational constraints) in the domains corresponding to each category within the Konect database, and where it was appropriate, the Network Repository as well \cite{Rossi2015}. See the Materials \& Methods section and SM Table \ref{table_RealNetworkData} for a full description of this selection process.

\begin{figure*}[t]
    \centering
    \subfloat[\label{fig:Figure4a}]{
        \includegraphics[width=0.95\textwidth]{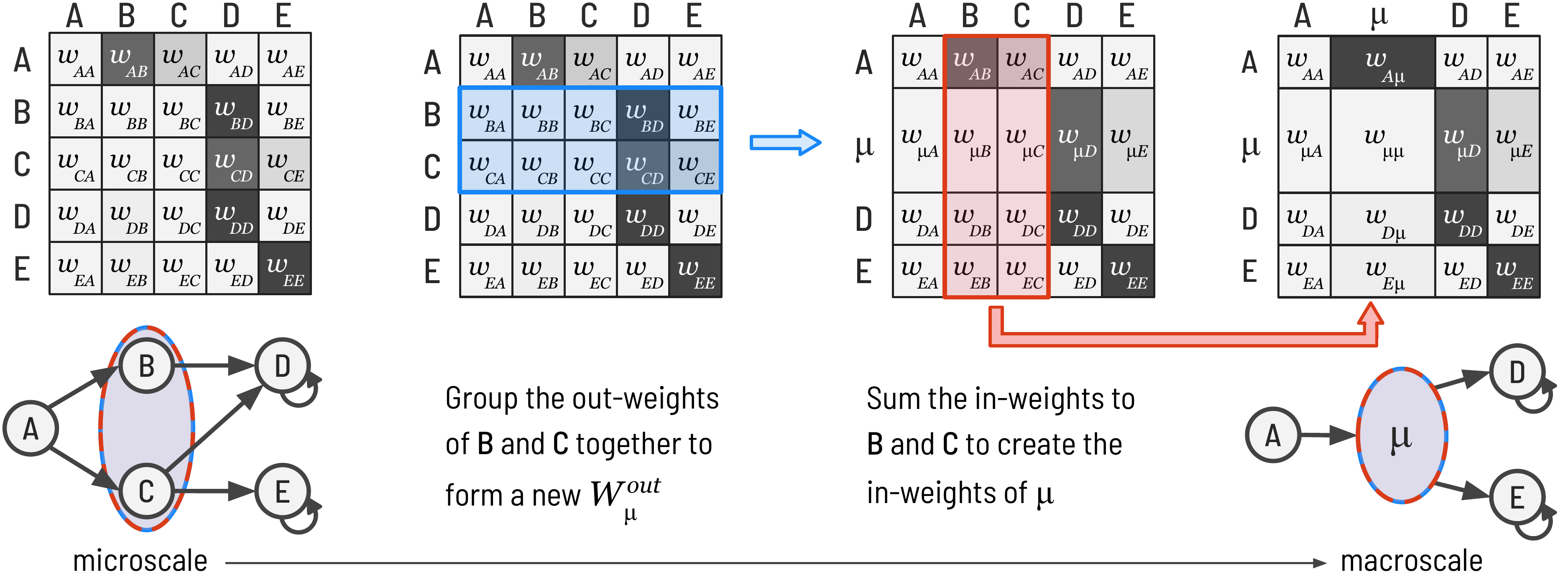}
    }\hfill
    \subfloat[\label{fig:Figure4b}]{
        \includegraphics[width=0.235\textwidth]{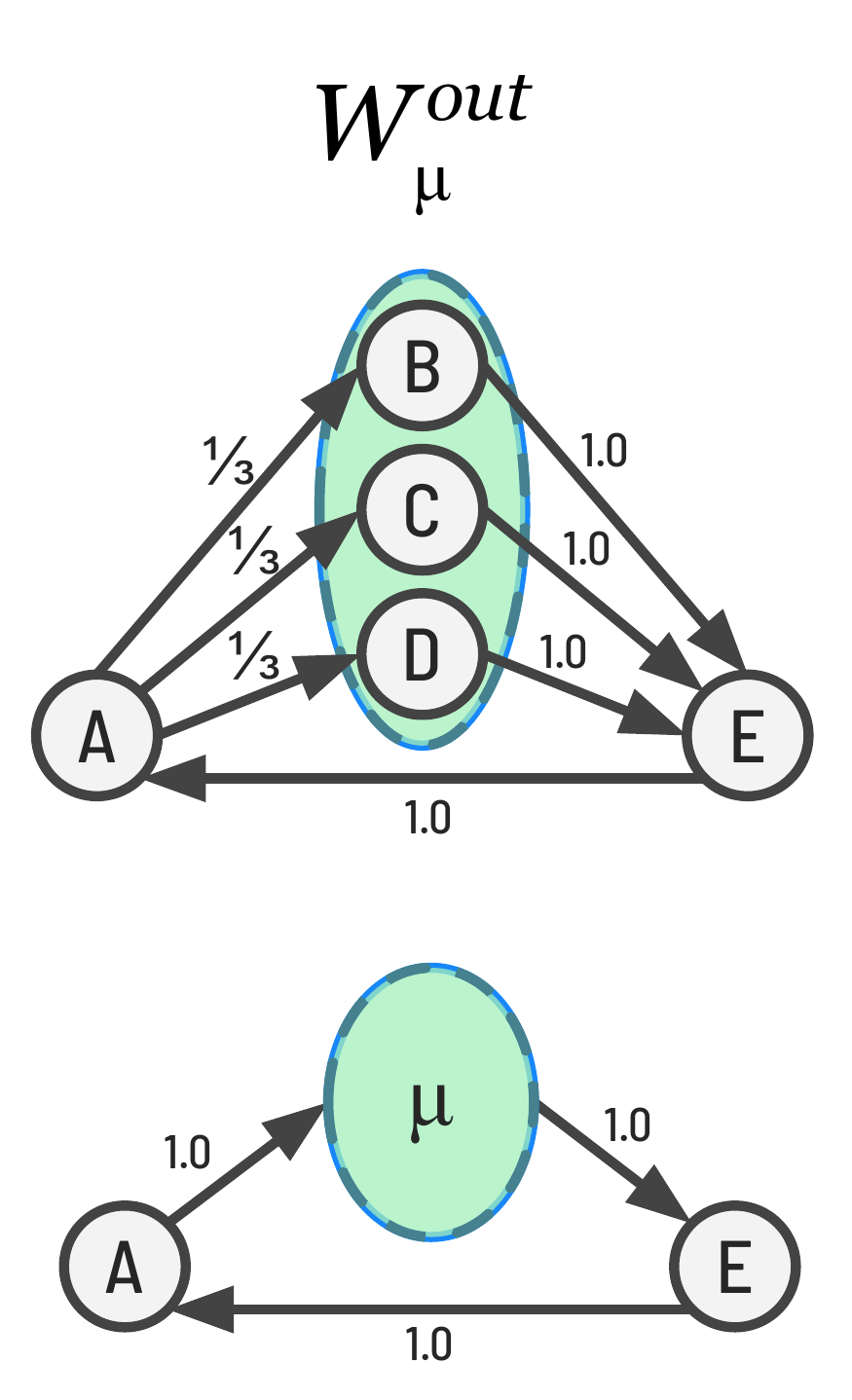}
    }
    \subfloat[\label{fig:Figure4c}]{
        \includegraphics[width=0.235\textwidth]{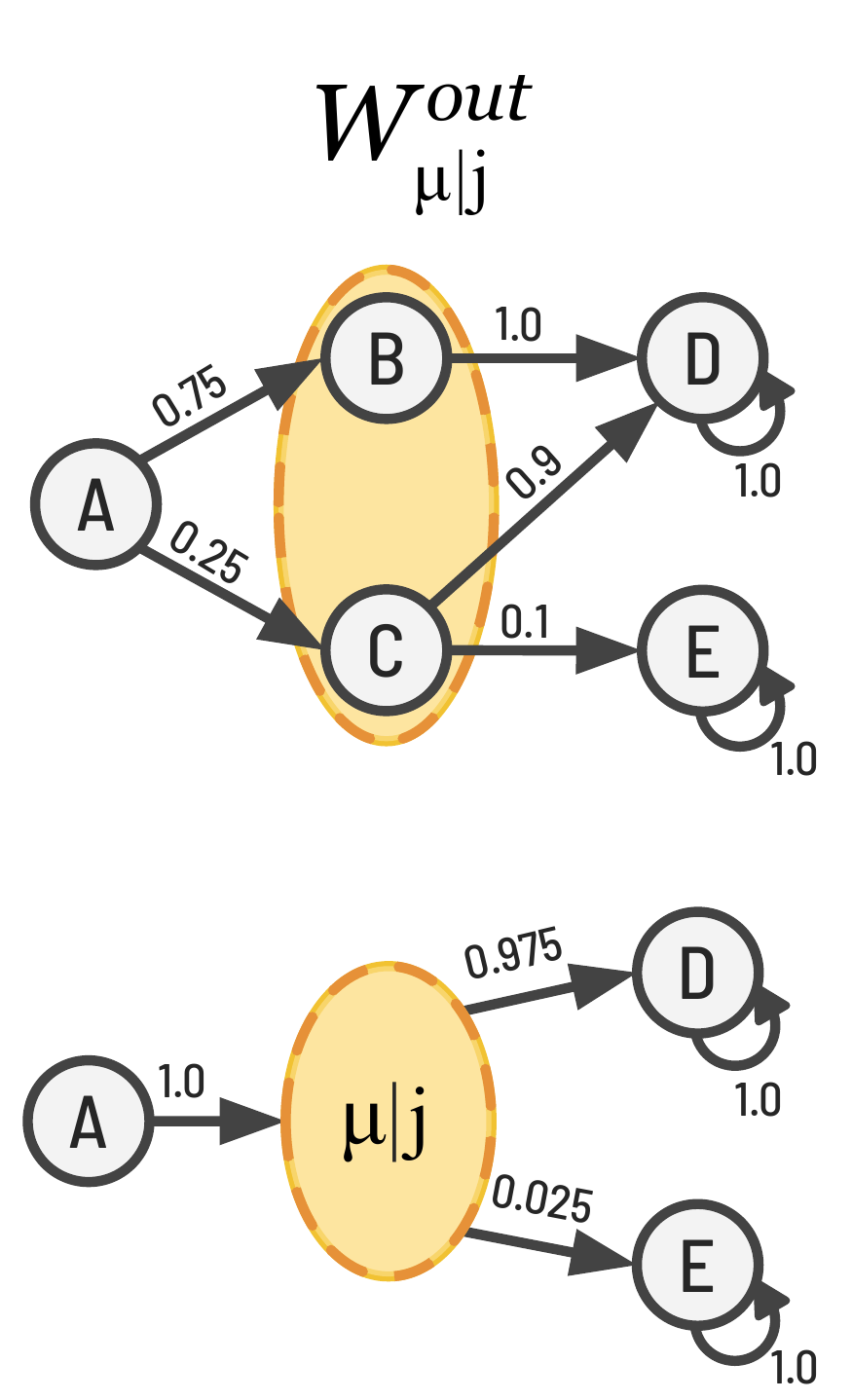}
    }
    \subfloat[\label{fig:Figure4d}]{
        \includegraphics[width=0.235\textwidth]{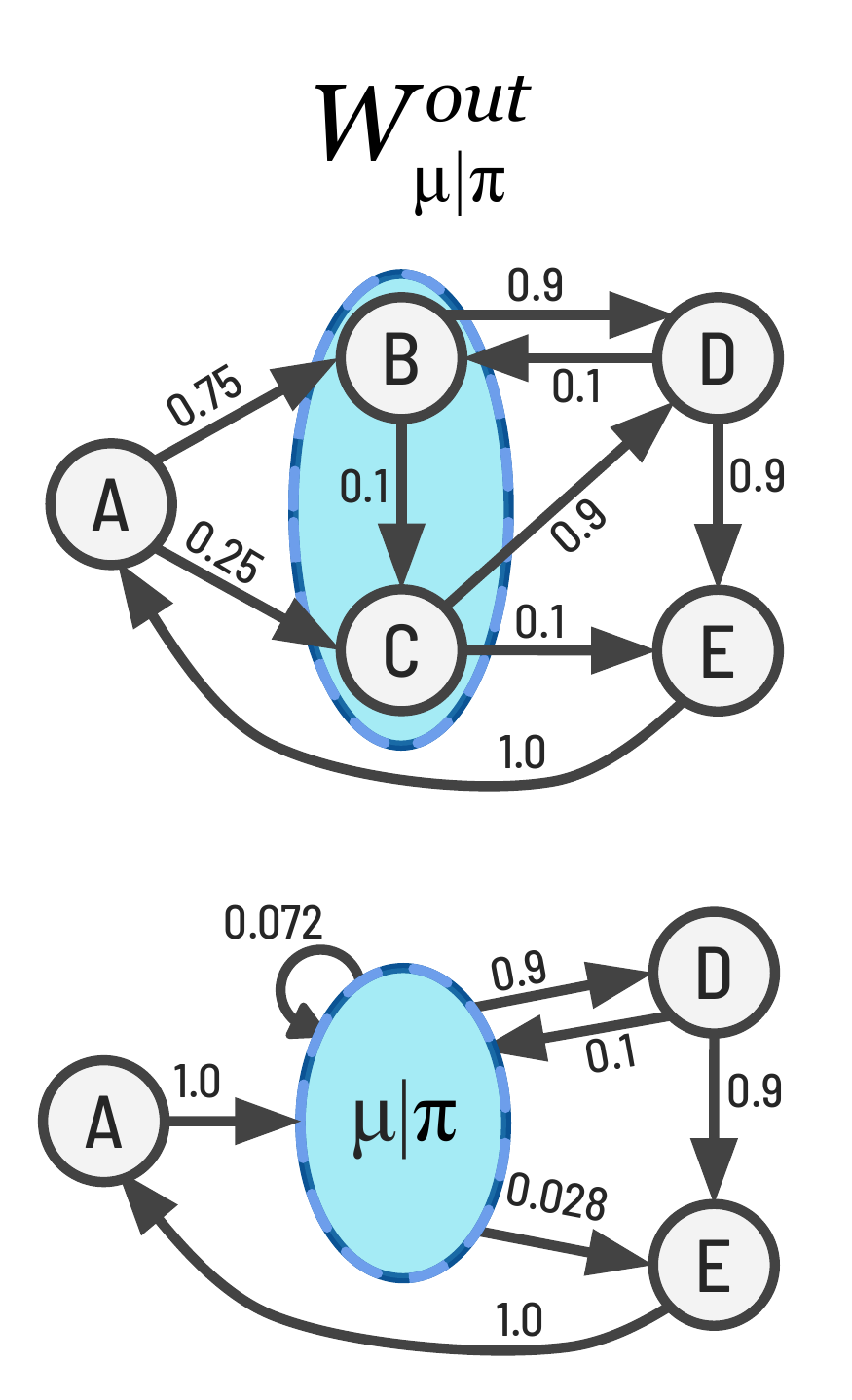}
    }
    \subfloat[\label{fig:Figure4e}]{
        \includegraphics[width=0.235\textwidth]{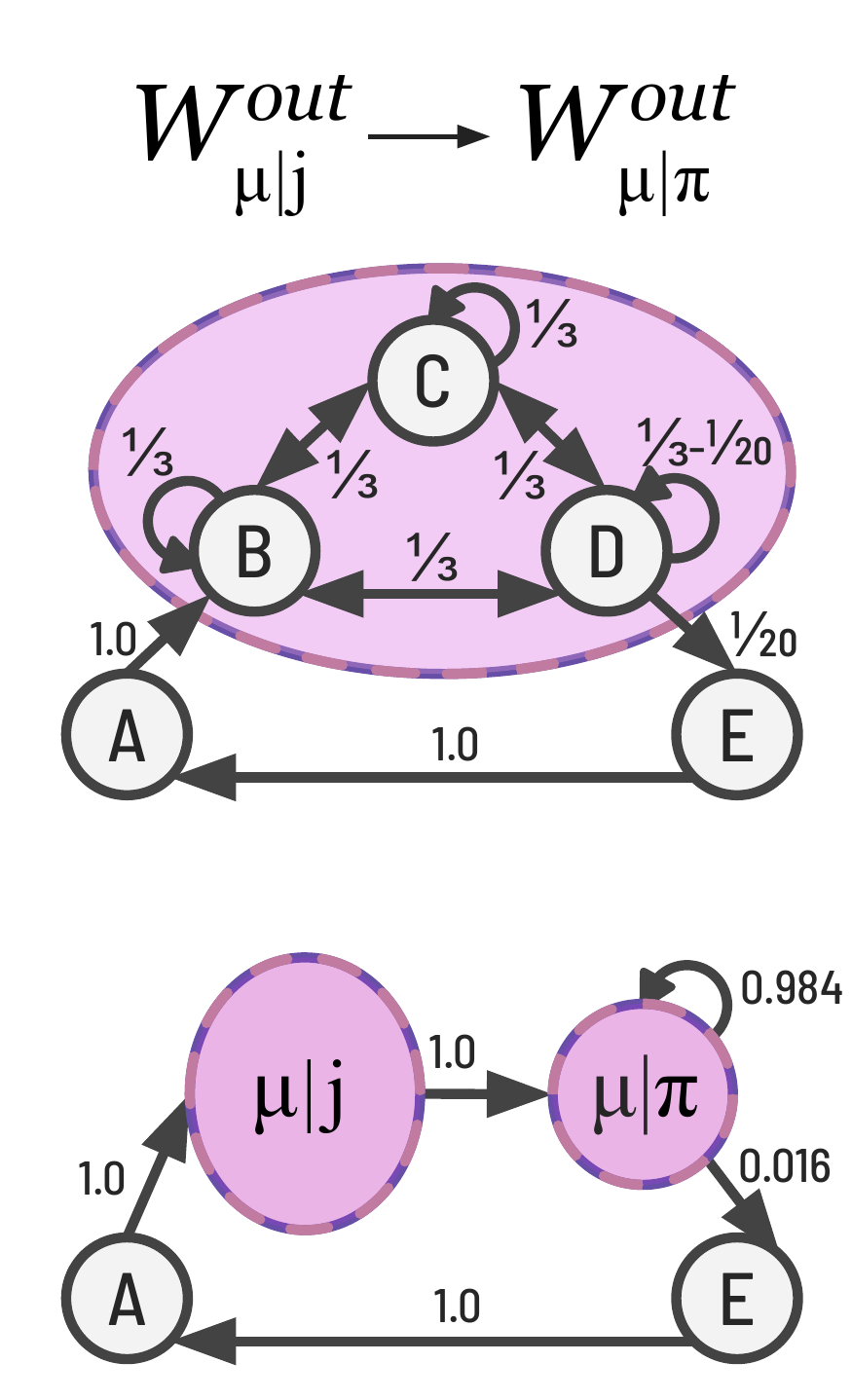}
    }
    \caption{\textbf{Macro-nodes.} (\textbf{A}) The original network, $G$ along with its adjacency matrix (left). The shaded oval indicates that subgraph $S$ member nodes $v_B$ and $v_C$ will be grouped together, forming a macro-node, ${\mu}$. All macro-nodes are some transformation of the original adjacency matrix via recasting it as a new adjacency matrix (right). The manner of this recasting depends on the type of macro-node. (\textbf{B}) The simplest form of a macro-node is when $W^{out}_{\mu}$ is an average of the $W^{out}_{i}$ of each node in the subgraph. (\textbf{C}) A macro-node that represents some path-dependency, such as input from $A$. Here, in averaging to create the $W^{out}_{\mu}$ the out-weights of nodes $v_B$ and $v_C$ are weighted by their input from  $v_A$. (\textbf{D}) A macro-node that represents the subgraph's output over the network's stationary dynamics. Each node has some associated ${\pi}_{i}$, which is the probability of ${v}_{i}$ in the stationary distribution of the network. The $W^{out}_{\mu}$ of a $\mu | \pi$ macro-node is created by weighting each $W^{out}_{i}$ of the micro-nodes in the subgraph $S$ by $\frac{{\pi}_{i}}{\sum_{k \in S} {\pi}_{k}}$. (\textbf{E}) A macro-node with a single timestep delay between input $\mu | j$ and its output $\mu | \pi$, each constructed using the same techniques as its components. However, $\mu | j$ always deterministically outputs to $\mu | \pi$. See SM \ref{sec:SI_keyTerms} for the full equations governing the creation of the $W^{out}_{\mu}$ of each of the different HOMs shown.}
    \label{fig:causal_emergence_example}
\end{figure*}

Lower effectiveness values correspond to structures that either have high degeneracy (as in right column, Fig. \ref{fig:Determ_Degen_hist}), low determinism (as in left column, Fig. \ref{fig:Determ_Degen_hist}), or a combination of both. In the networks we measured, biological networks on average have lower effectiveness values, whereas technological networks on average have the highest effectiveness. This finding aligns intuitively with what we know about the relationship between $EI$ and network structure, and it also supports long-standing hypotheses about the role of redundancy, degeneracy, and noise in biological systems \cite{Edelman2001, Tononi1999}. On the other hand, technological networks like power grids, autonomous systems, or airline networks on average are associated with higher effectiveness values. One explanation for this difference is that efficiency in human-made technological networks tends to create sparser, non-degenerate networks with higher effectiveness on average, wherein the nodes relationships are more specific in their targeting.

Perhaps it might be surprising to find that evolved networks have such low effectiveness. But, as we will show, a low effectiveness can actually indicate that there is informative higher-scale (macroscale) connectivity in the system. That is, a low effectiveness can reflect the fact that biological systems often contain higher-scale structure, which we demonstrate in the following section.

\subsection{\label{sec:causal_emergence}Causal emergence in complex networks}

This new global network measure, $EI$, offers a principled way to answer an important question: what is the scale that best captures the connectivity of a complex system? The resolution to this question is important because science analyzes the structure of different systems at different spatiotemporal scales, often preferring to intervene and observe systems at levels far above that of the microscale \cite{Hoel2018}. This is likely because relationships at the microscale can be extremely noisy and therefore uninformative, and coarse-graining can minimize this noise \cite{Hoel2013QuantifyingMicro}. Indeed, this noise minimization is actually grounded in Claude Shannon's noisy-channel coding theorem \cite{Shannon1948}, wherein dimension reductions can operate like codes that use more of a channel's capacity \cite{Hoel2017WhenTerritory}. Higher-level causal relationships often perform error-correction on the lower-level relationships, thus generating extra effective information at those higher scales. Measuring this difference provides a principled means of deciding when higher scales are more informative (emergence) or when higher scales are extraneous, epiphenomenal, or lossy (reduction).

Bringing these issues to network science, we can now ask: what representation will minimize the uncertainty present in a network? We do this by examining \textit{causal emergence}, which is is when a dimensionally-reduced network contains more informative connectivity, in the form of a higher $EI$ than the original network. Note that, as discussed, $EI$ can be interpreted solely as a general structural property of networks. Therefore, while we still call this phenomenon ``causal emergence'' because it has the same mathematical formalization as previous work in Boolean networks and Markov chains \cite{Hoel2013QuantifyingMicro, Hoel2017WhenTerritory, Hoel2018}, here we focus on how it can be used to identify the informative higher scales of networks regardless of what those networks represent.

Notably, the phenomenon can be measured by recasting networks at higher scales and observing how the $EI$ changes, a process which identifies whether the network's higher scales add information above and beyond lower scales.

\subsection{\label{sec:macro_scales}Network macroscales}

First we must introduce how to recast a network, $G$, at a higher scale. This is represented by a new network, $G_M$. Within $G_M$, a micro-node is a node that was present in the original $G$, whereas a macro-node is defined as a node, $\mu$, that represents a subgraph, $S_i$, from the original $G$ (replacing the subgraph within the network). Since the original network has been dimensionally reduced by grouping nodes together, $G_M$ will always have fewer nodes than $G$.

A macro-node $\mu$ is defined by some $W^{out}_{\mu}$, derived from the edge weights of the various nodes within the subgraph it represents. One can think of a macro-node as being a summary statistic of the underlying subgraph's behavior, a statistic that takes the form of a single node. Ultimately there are many ways of representing a subgraph, that is, building a macro-node, and some ways are more consistent than others in capturing the subgraph's behavior, depending on the connectivity. We highlight here that macroscales of networks should in general be \textit{consistent} with their underlying microscales in terms of their dynamics. While this has never been assessed within networks or systems generally, there has been previous research that has asked whether the macroscales of structural equation models are consistent with the effect of all possible interventions \cite{rubenstein2017causal}.

Here, to decide whether or not a macro-node is an consistent summary of its underlying subgraph, we formalize consistency as measure of whether random walkers behave identically on $G$ and $G_M$. We do this because random walks are often used to represent dynamics on networks \cite{Masuda2017}, and therefore many important analyses and algorithms---such as PageRank for determining a node's centrality \cite{Page1998} or InfoMap for community discovery \cite{Rosvall2008MapsStructure}---are based on random walks.

Specifically, we define the \textit{inconsistency} of a macroscale as the Kullback-Leibler divergence \cite{Cover2012} between the expected distribution of random walkers on $G$ vs. $G_M$, given some identical initial distribution on each. The expected distribution over $G$ at some future time, $t$, is $P_m(t)$, while the distribution over $G_M$ at some future time $t$ is $P_M(t)$. To compare the two, the distribution $P_m(t)$ is summed over the same nodes in the macroscale $G_M$, resulting in the distribution $P_{M|m}(t)$ (the microscale given the macroscale). We can then define the macroscale inconsistency over some series of time steps $T$ as:
\begin{equation}\label{eq:inaccuracy}
    \text{inconsistency} = \sum_{t=0}^T \text{D}_{_{KL}}[P_{M}(t) || P_{M|m}(t)] \tag{6}
\end{equation}

This consistency measure addresses the extent to which a random dynamical process on the microscale topology will be recapitulated on a dimensionally-reduced topology (for how this is applied in our analysis, see Materials \& Methods).

What constitutes a consistent macroscale depends on the connectivity of the subgraph that gets grouped into a macro-node, as shown in Fig. \ref{fig:causal_emergence_example}. The $W^{out}_{\mu}$ can be constructed based on the collective $W^{out}$ of the subgraph (shown in Fig. \ref{fig:Figure4a}). For instance, in some cases one could just coarse-grain a subgraph by using its average $W^{out}$ as the $W^{out}_{\mu}$ of some new macro-node $\mu$ (as in Fig. \ref{fig:Figure4b}). However, it may be that the subgraph has dependencies not captured by such a coarse-grain. Indeed, this is similar to the recent discovery that when constructing networks from data it is often necessary to explicitly model higher-order dependencies by using higher-order nodes so that the dynamics of random walks to stay true to the original data \cite{Xu2016}. We therefore introduce \textit{higher-order macro-nodes} (HOMs), which draw on similar techniques to consistently represent subgraphs as single nodes \cite{Xu2016}.

Different subgraph connectivities require different types of HOMs to consistently represent them. For instance, HOMs can be based on the input weights to the macro-node, which take the form $\mu | j$. In these cases the $W^{out}_{\mu|j}$ is a weighted average of each node's $W^{out}$ in the subgraph, where the weight is based on the input weight to each node in the subgraph (Fig. \ref{fig:Figure4c}). Another type of HOM that generally leads to consistent macro-nodes over time is when the $W^{out}_{\mu}$ is based on the stationary output from the subgraph to the rest of the network, which we represent as $\mu | \pi$ (Fig. \ref{fig:Figure4d}). These types of HOMs may have minor inconsistencies given some initial state, but will almost always trend toward perfect consistency as the network approaches its stationary dynamics (outlined in Section \ref{sec:MaterialsMethods}).

Subgraphs with complex internal dynamics can require a more complex type of HOM in order to preserve the macro-node's consistency. For instance, in cases where subgraphs have a delay between their inputs and outputs, this can be represented by a combination of $\mu | j$ and $\mu | \pi$, which when combined captures that delay (Fig. \ref{fig:Figure4e}). In these cases the macro-node $\mu$ has two components, one of which acts as a buffer over a timestep. This means that macro-nodes can possess memory even when constructed from networks that are at the microscale memoryless, and in fact this type of HOM is sometimes necessary to consistently capture the microscale dynamics.

We present these types of macro-nodes not as an exhaustive list of all possible HOMs, but rather as examples of how to construct higher scales in a network by representing subgraphs as nodes, and also sometimes using higher-order dependencies to ensure those nodes are consistent. This approach offers a complete generalization of previous work on coarse-grains \cite{Hoel2013QuantifyingMicro} and also black boxes \cite{ashby1957, Hoel2017WhenTerritory, Marshall2018}, while simultaneously solving the previously unresolved issue of macroscale consistency by using higher-order dependencies. The types of macro-nodes formed by subgraphs also provides substantive information about the network, such as whether the macroscale of a network possesses memory or path-dependency.

\subsection{\label{sec:causal_emergence_results}Causal emergence reveals the scale of networks}

A network has an informative macroscale when a recast network, $G_M$ (a macroscale), has more $EI$ than the original network, $G$ (the microscale). In general, networks with lower effectiveness (low $EI$ given their size) have a higher potential for such emergence, since they can be recast to reduce their uncertainty. Searching across groupings allows the identification or approximation of a macroscale that maximizes the $EI$.

Checking all possible groupings is computationally intractable for all but the smallest networks. Therefore, in order to find macro-nodes which increase the $EI$, we use a greedy algorithm that groups nodes together and checks if the $EI$ increases. By choosing a node and then pairing it iteratively with its surrounding nodes we can grow macro-nodes until pairings no longer increase the $EI$, and then move on to a new node (see the Materials \& Methods section for details on this algorithm).

\begin{figure*}[t!]
    \centering
    \subfloat[\label{fig:ce_pa}]{
        \includegraphics[width=1.0\columnwidth]{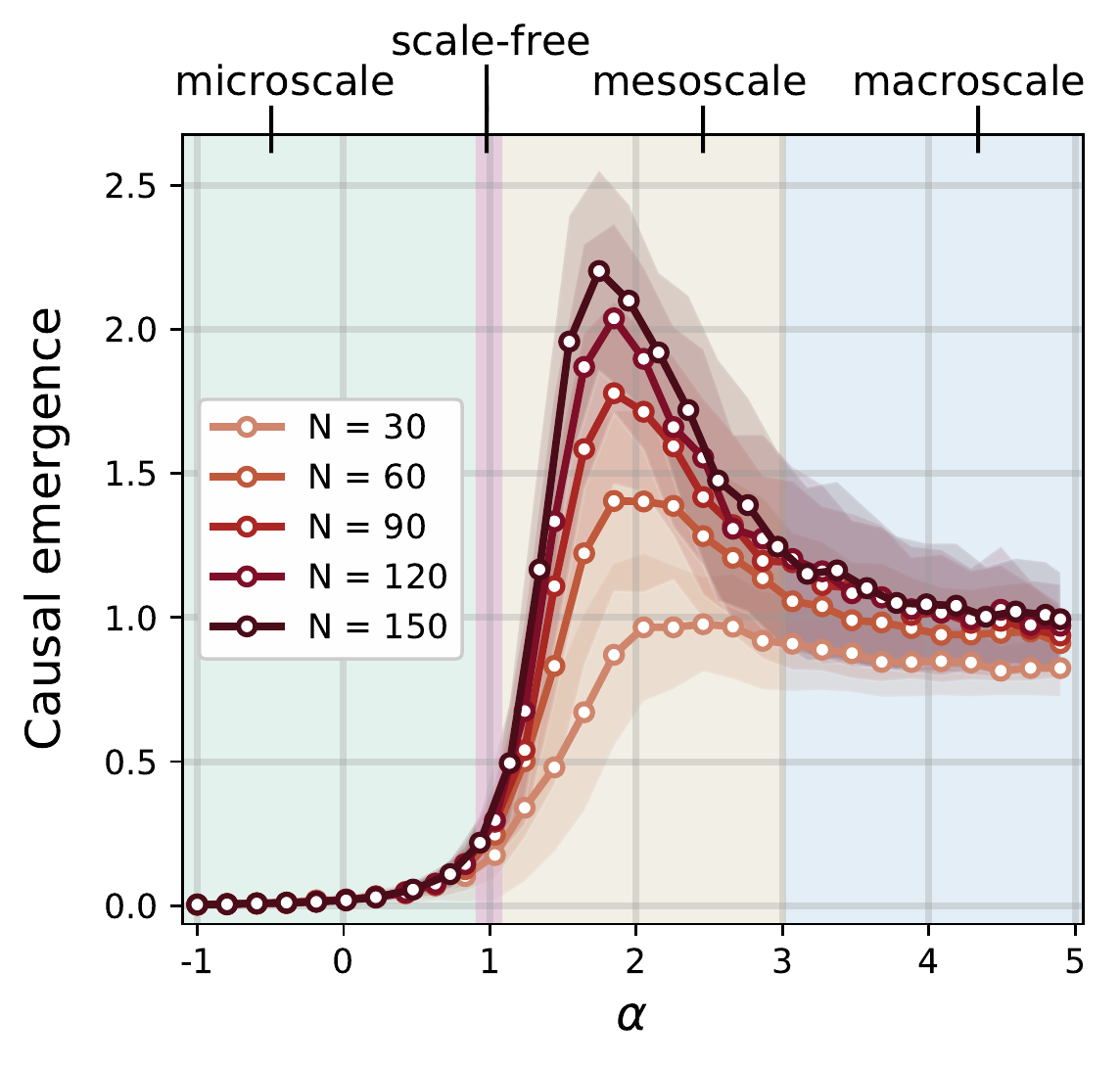}
    }\hfill
    \subfloat[\label{fig:ce_pa_size}]{
        \includegraphics[width=1.0\columnwidth]{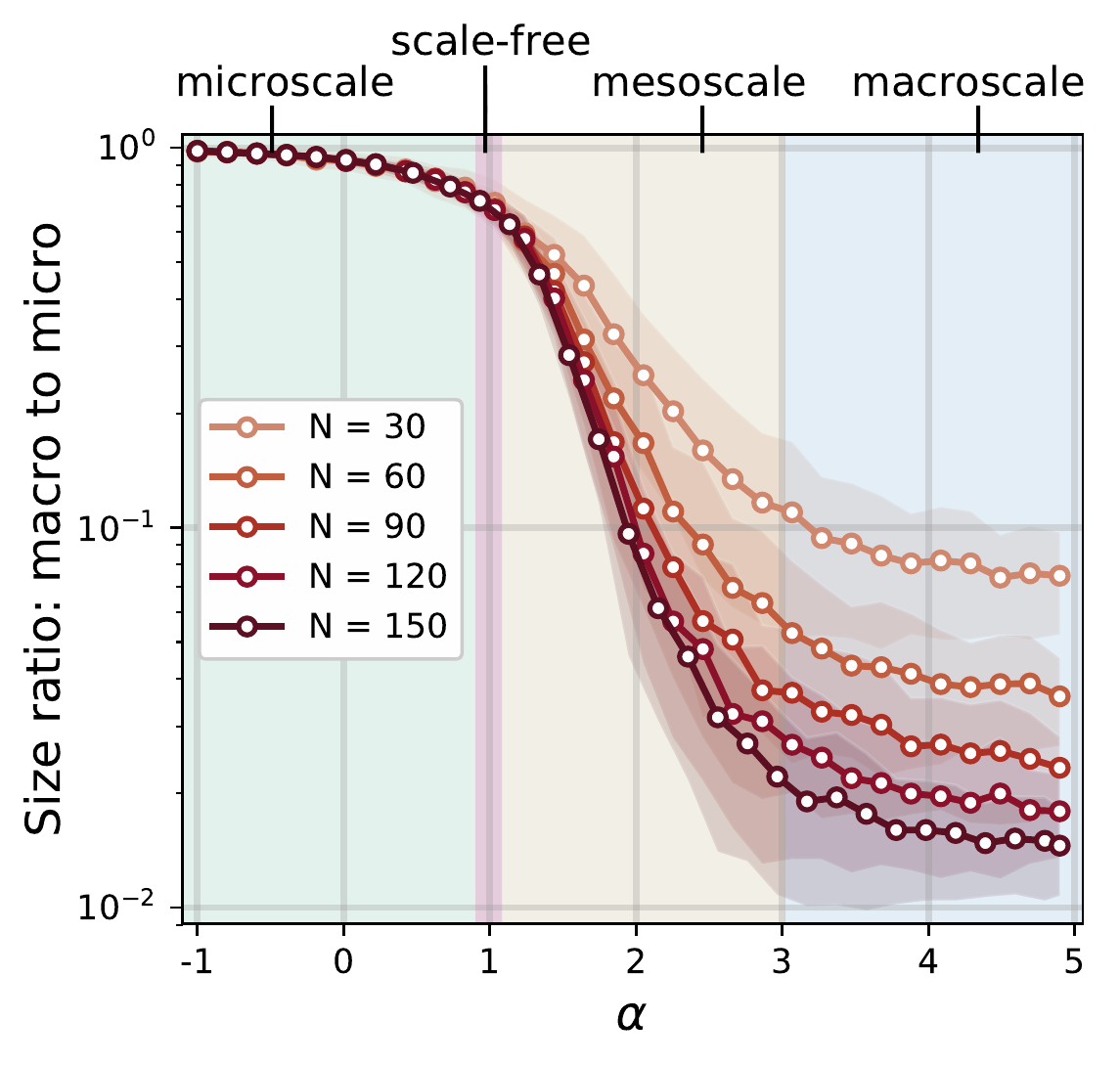}
    }
    \caption{\textbf{The emergence of scale in preferential attachment networks.} (\textbf{A}) By repeatedly simulating networks with different degrees of preferential attachment ($\alpha$ values) with $m=1$ new edge per each new node, and running them through a greedy algorithm (described in Materials \& Methods), we observe a distinctive peak of causal emergence once the degree of preferential attachment is above $\alpha=1$, yielding networks that are no longer ``scale-free.'' (\textbf{B}) The log of the ratio of original network size, $N$, to the size of the macroscale network, $N_M$. Networks with higher $\alpha$ values---more star-like networks---show drastic dimension reductions, and in fact all eventually reach the same $N_M$ of 2. Comparatively, random trees ($\alpha=0.0$) show essentially no informative dimension reductions.}
    \label{fig:ce_pa_alpha}
\end{figure*}

By generating undirected preferential attachment networks and varying the degree of preferential attachment, $\alpha$, we observe a crucial relationship between preferential attachment and causal emergence. One of the central results in network science has been the identification of ``scale-free'' networks \cite{Barabasi1999}. Our results show that networks that are not ``scale-free'' can be further separated into micro-, meso-, and macroscales depending on their connectivity. This scale can be identified based on their degree of causal emergence (Fig. \ref{fig:ce_pa}). In cases of sublinear preferential attachment ($\alpha < 1.0$) networks lack higher scales. Linear preferential attachment ($\alpha=1.0$) produces networks that are scale-free, which is the zone of preferential attachment right before the network develops higher scales. Such higher scales only exist in cases of superlinear preferential attachment ($\alpha > 1.0$). And past $\alpha > 3.0$ the network begins to converge to a macroscale where almost all the nodes are grouped into a single macro-node. The greatest amount of causal emergence is found in mesoscale networks, which is when $\alpha$ is between 1.5 and 3.0, when networks possess a rich array of macro-nodes. Note that the increase in $EI$ following macro-scale groupings for $\alpha > 1.0$ shown in Fig. \ref{fig:ce_pa} resembles the decrease in $EI$ with higher $\alpha$ that we observe in Fig. \ref{fig:ei_alpha}. This is because after $\alpha > 1.0$ the decreasing $EI$ of the microscale leaves room for improvement of the $EI$ at the macroscale, following a grouping of nodes.

Correspondingly the size of $G_M$ decreases as $\alpha$ increases and the network develops an informative higher scale, which can be seen in the ratio of macroscale network size, $N_M$, to the original network size, $N$ (Fig. \ref{fig:ce_pa_size}). As discussed previously, networks generated with higher values for $\alpha$ will be more and more star-like. Star-like networks have higher degeneracy and thus less $EI$, and because of this, we expect that there are more opportunities to increase the network's $EI$ through grouping nodes into macro-nodes. Indeed, the ideal grouping of a star network is when $N_M=2$ and $EI=1$ bit. This result is similar to recent advances in spectral coarse-graining that also observe that the ideal coarse-graining of a star network is to collapse it into a two-node network, grouping all the spokes into a single macro-node \cite{Laurence2019SpectralNetworks}, which is what happens to star networks that are recast as macroscales.

Our results offer a principled and general approach to such community detection by asking when there is an informational gain from replacing a subgraph with a single node. Therefore we can define \textit{causal communities} as being when a cluster of nodes, or some subgraph, forms a viable macro-node (note that this assumes the connections in the network actually represent possible causal interactions, but it also merely a topological property). Fundamentally causal communities represent noise at the microscale. The closer a subgraph is to complete noise, the greater the gain in $EI$ by replacing it with a macro-node (see SM \ref{sec:SI_characterize_CE}). Minimizing the noise in a given network also identifies the optimal scale to represent that network. However, there must be some structure that can be revealed by noise minimization in the first place. In cases of random networks that form a single large component which lacks any such structure, causal emergence does not occur (as shown in SM \ref{sec:SI_characterize_CE}).

\subsection{\label{sec:causal_emergence_realnetworks}Causal emergence in real networks}

\begin{figure}[t!]
    \centering
    \includegraphics[width=0.99\columnwidth]{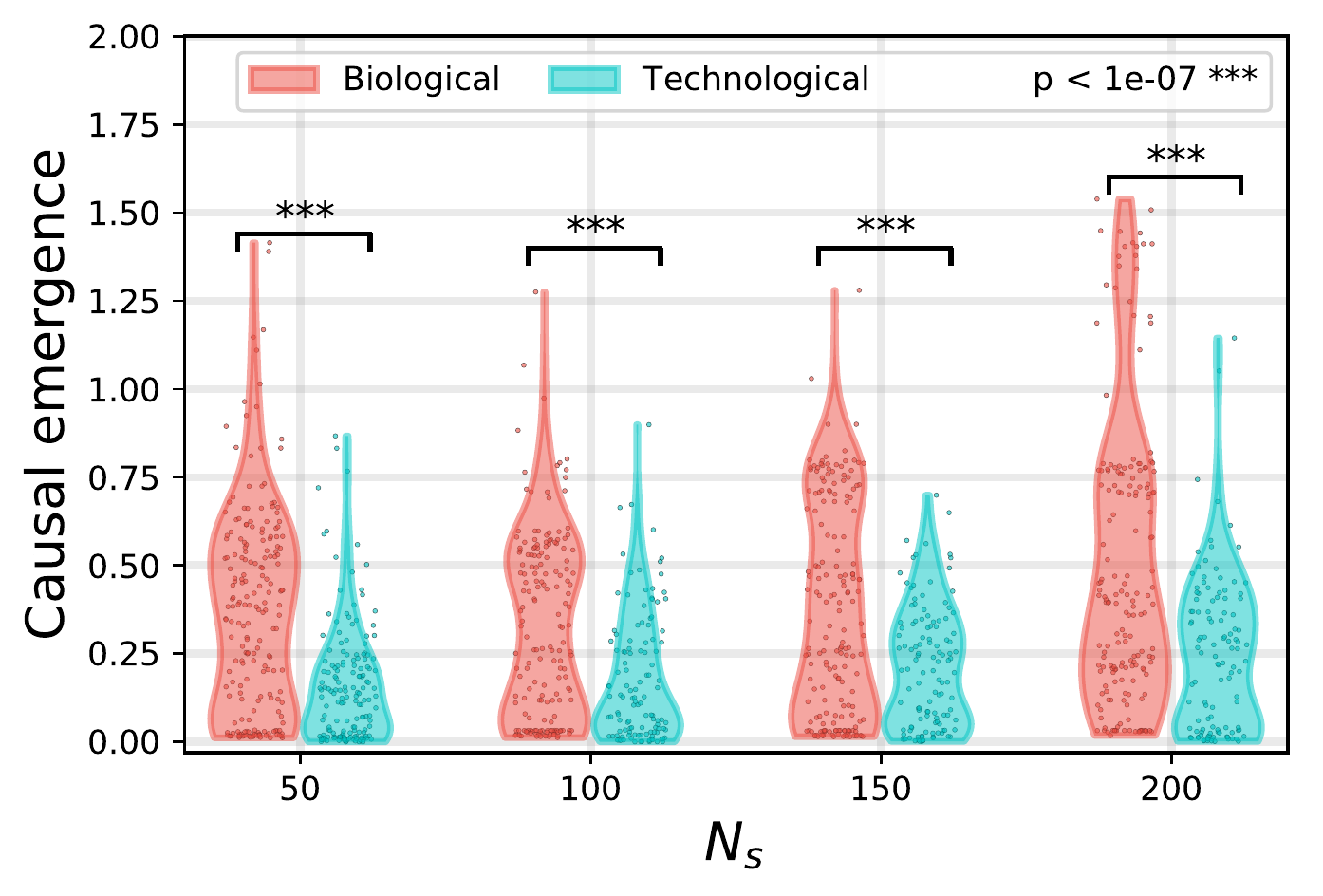}
    \caption{\textbf{Propensity for causal emergence in real networks.} Growing snowball samples of the two network domains that previously showed the greatest divergence in \textit{effectiveness}: technological and biological networks. At each snowball size, $N_s$, each network is sampled 20 times. Across these samples the total amount of causal emergence for a given sample size is significantly different between the two domains (t-test, comparison of means).}
    \label{fig:causal_emergence_konect}
\end{figure}

The presence and informativeness of macroscales should vary across real networks, dependent on connectivity. Here we investigate the disposition toward causal emergence of real networks across different domains. We draw from the same set of networks that are analyzed in Fig. \ref{fig:ei_real_nets}, the selection process and details of which is outlined in the Materials \& Methods section. The network sizes span up to 40,000 nodes, thus making it unfeasible to find the the best macroscales for each of them. Therefore, we focus specifically on the two categories that previously showed the greatest divergence in terms of the $EI$: biological and technological. Since we are interested in the general question of whether biological or technological networks show a greater disposition or propensity for causal emergence, we approximate causal emergence by calculating the causal emergence of sampled subgraphs of growing sizes. Each sample is found using a ``snowball sampling'' procedure, wherein a node is chosen randomly and then a weakly connected subgraph of a specified size is found around it \cite{Heckathorn2017}. This subgraph is then analyzed using the previously described greedy algorithmic approach to find macro-nodes that maximized the $EI$ in each network. Each available network is sampled 20 times for each size taken from it. In Fig. \ref{fig:causal_emergence_konect}, we show how the causal emergence of these real networks differentiates as we increase the sampled subgraph size, in a sequence of 50, 100, 150, and finally 200 nodes per sample. Networks of these sizes previously provided ample  evidence of causal emergence in simulated networks, as in Fig. \ref{fig:ce_pa}. Comparing the two categories of real networks, we observe a significantly greater propensity for causal emergence in biological networks, and that this is more articulated the larger the samples are. Note that constructing a random null model of these networks (e.g., a configuration model) would tend to create networks with minimal or negligible causal emergence, as is the case for ER networks (Fig. \ref{fig:CE_ER} in SM \ref{sec:SI_characterize_CE}).

That subsets of biological systems show a high disposition toward causal emergence is consistent, and even explanatory, of many long-standing hypotheses surrounding the existence of noise and degeneracy in biological systems \cite{Tononi1998ComplexityBrain}. It also explains the difficulty of understanding how the causal structure of biological systems function, since they are cryptic by containing certainty at one level and uncertainty at another.

\section{\label{sec:Discussion}Discussion}
We have shown that the information in the relationships between nodes in a network is a function of the uncertainty intrinsic to their connectivity, as well as how that uncertainty is distributed. To capture this information we adapted a measure, effective information ($EI$), for use in networks, and analyzed what it reveals about common network structures that have been studied by network scientists for decades. For example, the $EI$ of an ER random network tends to $-\log_2(p)$, and whether the $EI$ of a preferential attachment network grows or shrinks as new nodes are added is a function of whether its degree of preferential attachment, $\alpha$, is greater or less than 1.0. In networks where the mechanisms or transitions are unknown, but the structure is known, $EI$ captures the degree of unique targeting in the network. In real networks, we showed that the $EI$ of biological networks tends to be much lower than technological networks.

We also illustrated that what has been called ``causal emergence'' can occur in networks. This is the gain in $EI$ that occurs when a network, $G$, is recast as a new network, $G_M$. Finding this sort of informative higher scale means balancing the minimization of uncertainty while simultaneously maximizing the number of nodes in the network. These methods may be useful in improving scientific experimental design, the compression and search of big data, model choice, and even machine learning. Importantly, not every recast network, $G_M$, will have a higher $EI$ than the $G$ that it represents, that is, these same techniques can identify cases of reduction. Ultimately, this is because comparing the $EI$ of different network representations provides a ground for comparing the effectiveness of any two network representations of the same complex system. These techniques allow for the formal identification of the scale of a network. Scale-free networks can be thought of as possessing a fractal pattern of connectivity \cite{Barabasi2001}, and our results show that the scale of a network is the breaking of that fractal in one direction or the other.

The study of higher-order structures in networks is an increasingly rich area of research \cite{Rosvall2008MapsStructure, Benson2016, Scholtes2017, Yin2018, Lambiotte2019}, often focusing on constructing networks that better capture the data they represent. Here we introduce a formal and generalized way to recast networks at a higher scales while preserving random walk dynamics. In many cases, a macroscale of a network can be just as consistent in terms of random walk dynamics and also possess greater $EI$. Some macro-nodes in a macroscale may be of different types with different higher-order properties. In other words, we show how to turn a lower-order network into a higher-order network. One noteworthy and related aspect of our work is demonstrating how a system that is memoryless at the microscale can actually possess memory at the macroscale, indicating that whether a system has memory is a function of scale.

While some \cite{Stanley2018} have previously recast subgraphs as individual nodes as we do here, they have not done so in ways that are based on noise minimization and maximizing consistency, focusing instead on gains to algorithmic speed via compression. Explicitly creating macro-nodes to minimize noise brings the dependencies of the network into focus. This means that causal emergence in networks has a direct relationship to community detection, a vast sub-discipline that treats dense subgraphs within a network as representing shared properties, membership, or functions \cite{Fortunato2016, Radicchi2004}. However, the relationship between causal emergence and traditional community detection is not as direct as it may seem. For one, causal emergence is high in networks with high degeneracy (i.e. networks with high-degree hubs, as we show in Fig. \ref{fig:ce_pa}). Community detection algorithms do not typically select for such structural properties, instead focusing on dense subgraphs that connect more highly within the subgraph than outside \cite{Fortunato2016}. In SI Fig. \ref{fig:sbm_ce}, we show a landscape of stochastic block model networks and their associated values for causal emergence. Indeed in networks that would have high modularity \cite{Newman2004} (e.g. two disconnected cliques), we do observe causal emergence, but only when the two disconnected cliques are of \textit{different sizes}. This distinction is key and situates networks that display causal emergence in a meaningful place in the study of complex networks. In light of this, macro-nodes offer a sort of community detection where the micro-nodes that make up a macro-node are a community, and ultimately can be replaced by a macro-node that summarizes their behavior while reducing the subgraph's noise. Under this interpretation, community structure is characterized by noise rather than shared memberships.

\section{\label{sec:MaterialsMethods}Materials and Methods}

\subsection{\label{sec:SI_real_networks}Selection of real networks}

Networks were chosen to represent the four categories of interest: social, informational, biological, and technological (see SM Fig. \ref{fig:Konect_full}, where we detail the same information as in Fig. \ref{fig:ei_real_nets}, but also include the source of the network data in addition to the effectiveness value of each network). We used all the available networks under 40,000 nodes (due to computational constraints) within all the domains in the Konect database that reflected our categories of interest. For our social category we used the domains \textit{Human Contact}, \textit{Human Social}, \textit{Social}, and \textit{Communication}. For our information category we used the domains \textit{Citations}, \textit{Co-authorship}, \textit{Hyperlinks}, \textit{Lexical}, and \textit{Software}. For our biological category we used the domains \textit{Trophic} and \textit{Metabolic}. Due to overlaps between the Konect database and the Network Repository \cite{Rossi2015} in these domains, and the paucity of other biological data in the Konect database, we also included the \textit{Brains} domain and the \textit{Ecology} domain from the Network Repository to increase our sample size (again, all networks within these domains under 40,000 nodes were included). For our technological category, we used the domains \textit{Computer} and \textit{Infrastructure} from the Konect database. Again due to overlap between the Konect database and the Network Repository, we also included the \textit{Technological} and \textit{Power Networks} domains from the Network Repository. For a full table of the networks used in this study, along with their source and categorization, see Table \ref{table_RealNetworkData}.

\subsection{\label{sec:SI_accuracy}Creating consistent macro-nodes}

Previously we outlined methods for creating consistent macro-nodes of different types. Here we explore their implementation, which requires deciding which macroscales are consistent. Inconsistency is measured as the Kullback-Leibler divergence between the expected distribution of random walkers on both the microscale ($G$) and the macroscale ($G_M$), given an initial distribution, as in Eq. \ref{eq:inaccuracy}.

To measure the inconsistency we use an initial maximum entropy distribution on the shared nodes between $G$ and $G_M$. That is, only the set of nodes that are left ungrouped in $G_M$. Similarly, we only analyze the expected distribution over that same set of micro-nodes. Since such distributions are only over a portion of the network, to normalize each distribution to 1.0 we include a single probability that represents all the non-shared nodes between $G$ and $G_M$ (representing when a random walker is on a macro-node).

We focus on the shared nodes between $G$ and $G_M$ for the inconsistency measure because: a) it is easy to calculate which is necessary during an algorithmic search, b) except for unusual circumstances the inconsistency over the shared nodes still reflects the network as a whole, and c) even in cases of the most extreme macroscales (such as when $\alpha > 4$ in Fig. \ref{fig:ce_pa_alpha}), there are still nodes shared between $G$ and $G_M$.

Here we examine our methods of using higher-order dependencies in order to demonstrate that this creates consistent macro-nodes. We use 1000 simulated preferential attachment networks, which were chosen as a uniform random sample between parameters $\alpha = 1.0$ and $2.0$, $n = 25$ to $35$, and with either $m=1$ or $2$. These networks were then grouped via the algorithm described in the following section. All macro-nodes were of the $\mu | \pi$ type and their inconsistency was checked over 1000 timesteps. These macro-nodes generally have consistent dynamics, either because they start that way or because they trend to that over time, and of the 1000 networks, only 4 had any divergence greater than 0 after 1000 timesteps. In Fig. \ref{fig:15inaccuracies} in SM \ref{sec:SI_inacc_examples}, we show 15 of these simulated networks, along with their parameters, number of macro-nodes, and consistencies. Note that even in the cases with early nonzero inconsistency, this is always very low in absolute terms of bits, and of the randomly chosen 15 none do not trend toward consistency over time. In our observations most macro-nodes converge before 500 timesteps, so therefore, in analyzing the real world networks using the $\mu | \pi$ macro-node we check all macro-nodes for consistency and only reject those that are inconsistent at 500 timesteps. More details about the algorithmic approach to finding causal emergence can be found in the following section.

\subsection{\label{sec:SI_greedyalgorithm}Greedy algorithm for causal emergence}

The greedy algorithm used for finding causal emergence in networks is structured as follows: for each node, $v_i$, in the shuffled node list of the original network, collect a list of neighboring nodes, $\{v_j\} \in B_i$, where $B_i$ is the \textit{Markov blanket} of $v_i$ (in graphical models, the Markov blanket, $B_i$, of a node, $v_i$, corresponds to the ``parents'', the ``children'', and the ``parents of the children'' of $v_i$ \cite{Friston2013LifeIt.}). This means that $\{v_j\} \in B_i$ consists of nodes with outgoing edges leading into $v_i$, nodes that the outgoing edges from $v_i$ lead into, and nodes that have outgoing edges leading into the out-neighbors of $v_i$. For each node in $\{v_j\}$, the algorithm calculates the $EI$ of a macroscale network after $v_i$ and $v_j$ are combined into a macro-node, $v_M$, according to one of the macro-node types in Fig. \ref{fig:causal_emergence_example}. If the resulting network has a higher $EI$ value, the algorithm stores this structural change and, if necessary, supplements the queue of nodes, $\{v_j\}$, with any new neighboring nodes from $v_j$'s Markov blanket that were not already in $\{v_j\}$. If a node, $v_j$, has already been combined into a macro-node via a grouping with a previous node, $v_i$, then it will not be included in new queues, $\{v_j'\}$, of later nodes to check. The algorithm iteratively combines such pairs of nodes until every node, $v_j$, in every node, $v_i$'s Markov blanket is tested.

\section*{Additional information} 
\noindent \textbf{Author contributions:} B.K. and E.H. conceived the project. B.K. and E.H. wrote the article. B.K. performed the analyses. \textbf{Acknowledgements:} The authors thank Conor Heins, Harrison Hartle, and Alessandro Vespignani for their insights about notation and formalism of effective information. \textbf{Funding:} This research was supported by the Allen Discovery Center program through The Paul G. Allen Frontiers Group (12171). This publication was made possible through the support of a grant from Templeton World Charity Foundation, Inc. (TWCFG0273). The opinions expressed in this publication are those of the author(s) and do not necessarily reflect the views of Templeton World Charity Foundation, Inc. This work was also supported in part by the National Defense Science \& Engineering Graduate Fellowship (NDSEG) Program. \textbf{Competing interests:} The authors declare no competing interests. \textbf{Data and software availability:} All data used in this work was retrieved from the Konect Database \cite{Kunegis2013} and also the Network Repository \cite{Rossi2015}, which are publicly available. Software for calculating $EI$ in networks and for finding causal emergence in networks is available by request or at \url{https://github.com/jkbren/einet}.

% \bibliography{main} 

%apsrev4-2.bst 2018-12-27 (MD) hand-edited version of apsrev4-1.bst

% \vfill
\clearpage
\pagebreak

\section{Supplementary Materials}

\subsection{\label{sec:SI_keyTerms}Table of key terms}

A table of key terms can be found in Table \ref{table_of_terms}.

\begin{table*}[t]
    \renewcommand*{\arraystretch}{1.75}
    \centering
    \begin{tabular}{| p{3.75cm} | p{6.25cm} | p{5.5cm} |}
        \hline
        \textbf{Term} & \textbf{Description} & \textbf{Notation} \\
        \hline \hline
        Network size & the number of nodes in the network & $ N $ \\ \hline
        Out-weight vector\newline($v_i$) & a vector of probabilities $w_{ij}$ that a random walker on node $v_i$ will transition to $v_j$ & $ W^{out}_{i} = \{w_{i1}, w_{i2},,... w_{ij}, ... w_{iN}\}$ \\ \hline
        Effective information\newline (network) & the total information in a causal structure, in bits & $EI = {{H}(\langle W^{out}_{i}\rangle)} - {\langle {H}(W^{out}_{i}) \rangle}$ \\ \hline
        Determinism\newline ($v_i$) & how certain about next steps is a random walker on $v_i$ & $\text{det}_{i} = \log_2(N) - H(W^{out}_{i})$ \\ \hline
        Degeneracy\newline (network) & how distributed the certainty is over the nodes of the network & $\text{degeneracy} = \log_2(N) - H(\langle W^{out}_{i}\rangle)$ \\ \hline
        Effect information\newline ($v_i$) & the contribution of each node $v_i$ to the network's $EI$ & $EI_{i} = \text{D}_{_{KL}}[ W^{out}_{i} || \langle W^{out}_{i}\rangle ]$ \\ \hline
        Micro-nodes in a\newline macro-node & the set of micro-nodes grouped into a macro-node in the new network, $G_M$ & $S = \{v_i, v_j, ...\}$, of length $N_S$ \\ \hline
        Macro-node out-weights & the out-weights from macro-node, $\mu$, to its neighbors & $W_{\mu}^{out} = \displaystyle\sum_{i \in S} W_i^{out} \cdot \Big(\dfrac{1}{N_S}\Big)$ \\ \hline
        Macro-node out-weights\newline given input weights & the out-weights from macro-node, $\mu$, to its neighbors, \textit{conditioned} on in-weights to the micro-nodes, $v_i \in S$ & \vspace{0.001cm} $W_{\mu|j}^{out} = \displaystyle\sum_{i \in S} W_i^{out} \cdot \Big(\dfrac{\sum_{j->i} w_{ji}}{\sum_{j->k \in S} w_{jk}}\Big)$ \\ \hline
        Macro-node out-weights\newline given the stationary\newline distribution & the out-weights from macro-node, $\mu$, to its neighbors, conditioned on the stationary probabilities, $\pi_i$, of micro-nodes, $v_i \in S$ & \vspace{0.01cm} $W_{\mu|\pi}^{out} = \displaystyle\sum_{i \in S} W_i^{out} \cdot \Big(\dfrac{\pi_i}{\sum_{k\in S} \pi_k}\Big)$ \\ \hline
    \end{tabular}
    \caption{\textbf{Table of key terms}. Quantities needed in order to calculate $EI$ and create consistent macro-nodes.}
    \label{table_of_terms}
\end{table*}

\subsection{\label{sec:ei_calc}Effective information calculation}

Mathematically, $EI$ has been expressed in a number of previous ways. The first was as the mutual information between two subsets of a system (while injecting noise into one), originally proposed as a step in the calculation of integrated information between neuron-like elements \cite{Tononi2004, Balduzzi2009}. More recently, it was pointed out that in general an intervention distribution, $I_D$, defined as a probability distribution over the $do(x)$ operator (as in \cite{Pearl2000}), creates some resultant effect distribution, $E_D$. Then the $EI$ is the mutual information, $I(I_D; E_D)$, between the two, when the interventions are done like a randomized trial to reveal the dependencies (i.e., at maximum entropy \cite{Fisher1936, Hoel2017WhenTerritory}). 

\begin{figure*}[t!]
    \centering
    \subfloat[\label{fig:example_network}]{
        \includegraphics[width=0.95\columnwidth]{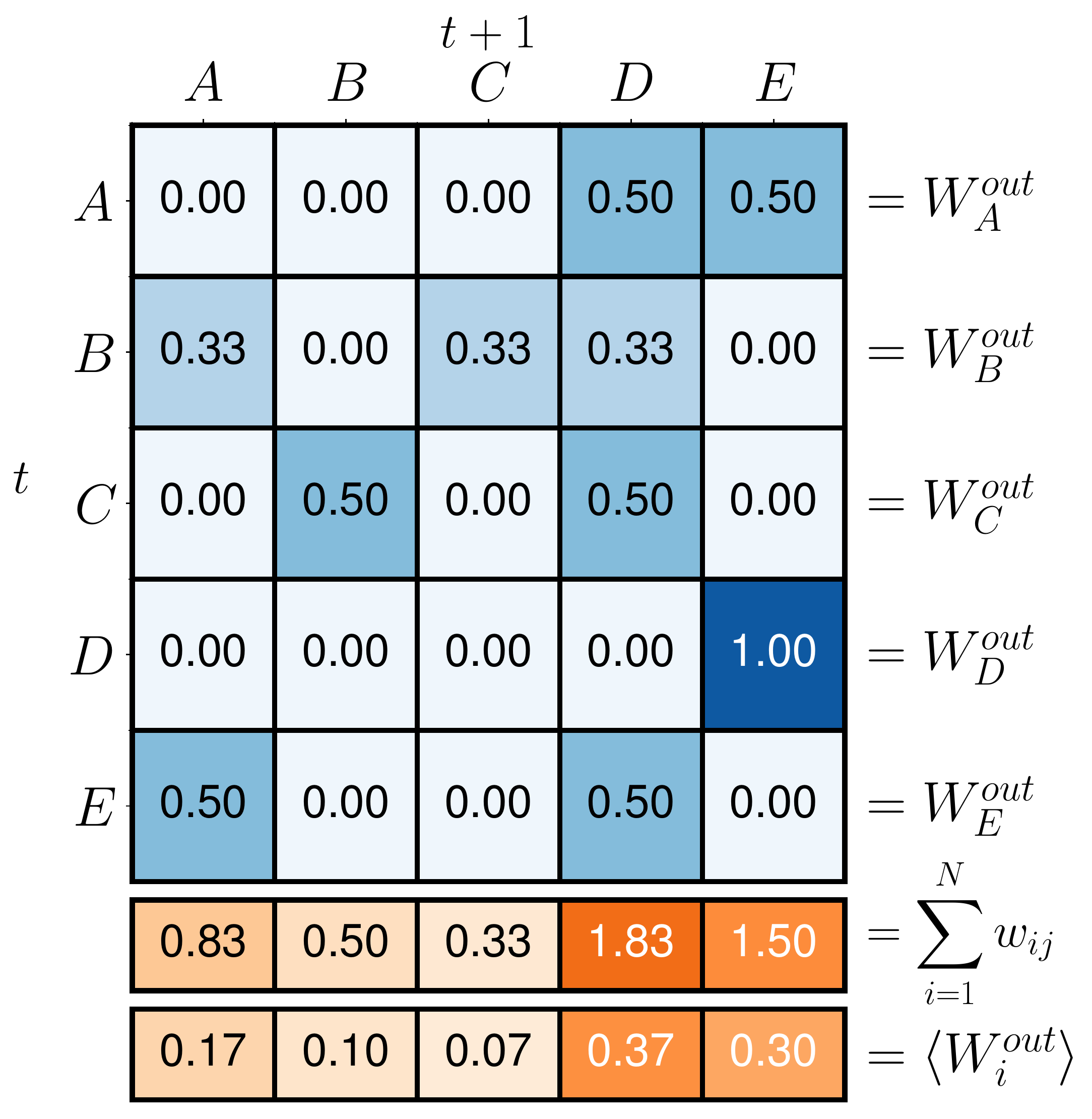}
    }\hfill
    \subfloat[\label{fig:example_kld}]{
        \includegraphics[width=\textwidth]{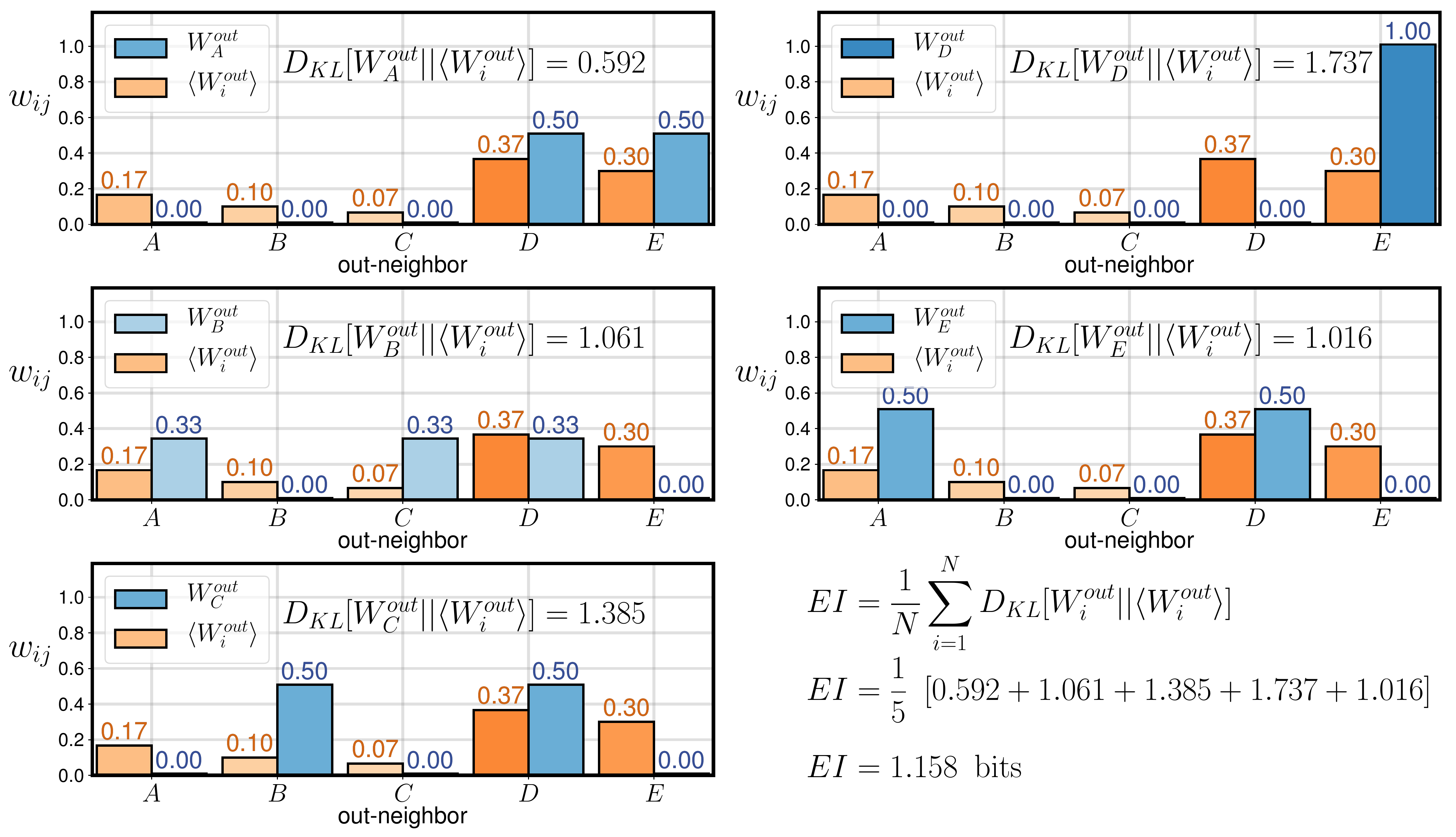}
    }
    \caption{\textbf{Illustration of the calculation of effective information.} (\textbf{A}) The adjacency matrix of a network with 1.158 bits of $EI$ (calculation shown in (\textbf{B})). The rows correspond to $W^{out}_{i}$, a vector of probabilities that a random walker on node $v_i$ at time $t$ transitions to $v_j$ in the following time step, $t+1$. $\langle W^{out}_{i}\rangle$ represents the (normalized) input weight distribution of the network, that is, the probabilities that a random walker will arrive at a given node $v_j$ at $t+1$, after a uniform introduction of random walkers into the network at $t$. (\textbf{B}) Each node's contribution to the $EI$ ($EI_i$) is the KL divergence of its $W^{out}_{i}$ vector from the network's $\langle W^{out}_{i}\rangle$, known as the \textit{effect information}.}
    \label{fig:example_calc}
\end{figure*}

In order to calculate the total information contained in the causal relationships of a system, $EI$ is applied to the system as a whole \cite{Hoel2013QuantifyingMicro}. There, $EI$ was defined over the set of all states of a system and its state transitions. Because the adjacency matrix of a network can be cast as a transition matrix (as in Fig. \ref{fig:example_network}), the $EI$ of a network can be expressed as:
\begin{equation}\label{eq:dkl_ei}
    EI = \dfrac{1}{N}   \sum_{i=1}^N \text{D}_{_{KL}}[W^{out}_{i} || \langle W^{out}_{i} \rangle] \tag{7}
\end{equation}

\noindent where $EI$ is the average of the \textit{effect information}, $EI_i$, of each node (see Table \ref{table_of_terms} and Fig. \ref{fig:example_kld}). This is equivalent to our derivation of $EI$ from first principles in Equation \ref{eq:ei}, since:
\begin{align*}
    EI &= \dfrac{1}{N} \sum_{i=1}^N \text{D}_{_{KL}}[W^{out}_{i} || {\langle W^{out}_{i}\rangle}] \\
    &= \dfrac{1}{N} \sum_{i=1}^{N} \sum_{j=1}^{N} w_{ij}\log_2\bigg(\dfrac{w_{ij}}{W_{j}}\bigg) \nonumber \\
    &= \dfrac{1}{N}  \sum_{i=1}^{N}\bigg( \sum_{j=1}^{N} w_{ij}\log_2(w_{ij}) - \sum_{j=1}^{N} w_{ij}\log_2(W_{j})\bigg) \nonumber \\
    &= \dfrac{1}{N}  \sum_{i=1}^{N} \sum_{j=1}^{N} w_{ij}\log_2\big(w_{ij}\big) - \dfrac{1}{N}  \sum_{i=1}^{N} \sum_{j=1}^{N} w_{ij}\log_2\big(W_{j}\big) \tag{8}\label{eq:decompose_ei}
\end{align*}

Note that for a given node, $v_i$, the term in the first summation in Equation \ref{eq:decompose_ei} above, $\displaystyle\sum_{j=1}^{N} w_{ij}\log_2\big(w_{ij}\big)$, is equivalent to the negative entropy of the out-weights from $v_i$, $-H(W_i^{out})$. Also note that $W_j$, the \textit{j}th element in the $\langle W^{out}_{i}\rangle$ vector, is the normalized sum of the incoming weights to $v_j$ from its neighbors, $v_i$, such that $W_j=\dfrac{1}{N} \displaystyle\sum_{i=1}^N w_{ij}$. We substitute these two terms into Equation \ref{eq:decompose_ei} above such that: 
\begin{equation*}
    EI = \dfrac{1}{N}   \sum_{i=1}^{N}-H(W_i^{out}) -  \sum_{j=1}^{N} W_j\log_2\big(W_{j}\big) \tag{9}
\end{equation*}

This is equivalent to the formulation of $EI$ from Equation \ref{eq:ei}, since $H(\langle W^{out}_{i}\rangle) = -\displaystyle\sum_{j=1}^{N} W_j\log_2(W_{j})$:
\begin{equation*}
    EI = H(\langle W^{out}_{i}\rangle) -\langle H(W_i^{out}) \rangle \tag{1}
\end{equation*}

In the  derivations of SM \ref{sec:SI_derivations} we adopt the relative entropy formulation of $EI$ from Equation \ref{eq:dkl_ei} for ease of derivation. For a visual intuition behind the calculations involved in this formulation of $EI$, see how the network in Fig. \ref{fig:example_network} is used to calculate its $EI$ (Fig. \ref{fig:example_kld}), by calculating the mean effect information, $EI_i$, of nodes in the network.

\subsection{\label{sec:SI_derivations}Deriving the effective information of common network structures} 

Here we inspect the $EI$ for iconic graphical structures, and in doing so, we see several interesting relationships between a network structure and its $EI$. First, however, we will introduce key terminology and assumptions. 

Let $\langle k \rangle$ be the average degree of a network, $G$, and each node, $v_i$, has degree, $k_i$. In directed graphs each $v_i$ has an in-degree, $k^{in}_{i}$, and an out-degree, $k^{out}_{i}$. These correspond to the number of edges leading in to $v_i$ and edges going out from $v_i$. The total number of edges in $G$ is represented by $E$. In undirected Erd\H{o}s-R\'enyi (ER) networks, the total number of edges is given by $E = p\frac{N(N-1)}{2}$, where $p$ represents the probability that any two nodes, $v_i$ and $v_j$, will be connected. In the following subsections, we derive the $EI$ of several prototypical network structures, from random graphs to ring lattices to star networks. Note that for the following derivations we proceed from the relative entropy formalism from SM \ref{sec:ei_calc}, and note that therefore $N$ is the number of nodes with the output, $N=N_{out}$.

\begin{figure*}[t]
    \centering
    \includegraphics[width=0.8\textwidth]{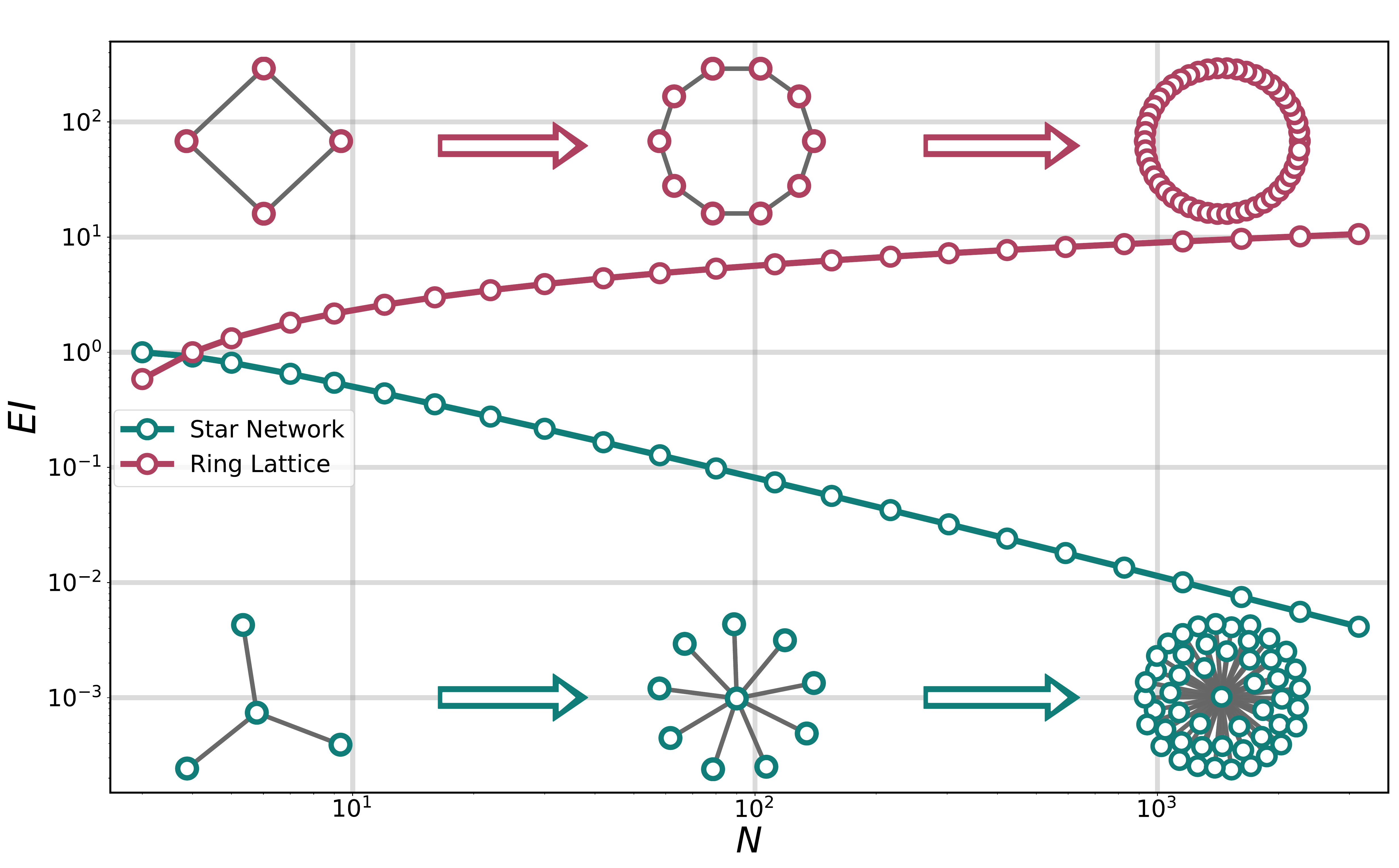}
    \caption{\textbf{Effective information of stars and rings.} As the number of nodes in star networks increases, the $EI$ approaches zero, while the $EI$ of ring lattice networks grows logarithmically as the number of nodes increases.}
    \label{fig:star_ring}
\end{figure*}

\subsubsection{\label{sec:SI_derivations_ER}Derivation: effective information of ER networks}

In Erd\H{o}s-R\'enyi networks, $EI$ does not depend on the number of nodes in the network, $N$. Instead, the network's $EI$ reaches its maximum at $-\log_2(p)$. This is because in ER networks, each node is expected to connect to $\langle k \rangle = pN$ neighboring nodes, such that every value in $W^{out}_{i} = \frac{1}{\langle k \rangle}$ and every value in $ \langle W^{out}_{i}\rangle = \frac{\langle k \rangle}{N\langle k \rangle}=\frac{1}{N}$, which can be represented as:
\begin{align*}
    EI_{ER} &= \dfrac{1}{N}   \sum_{i}^{N} \text{D}_{_{KL}}\Big[\big(\dfrac{1}{\langle k \rangle},\dfrac{1}{\langle k \rangle}, ...\big) || \big(\dfrac{1}{N}, \dfrac{1}{N}, ...\big)\Big] 
\end{align*}

Each node in an ER network is expected to be identical to all other nodes in the network, and calculating the expected \textit{effect information}, $EI_i$, is equivalent to calculating the network's $EI$. As such, we observe: 
\begin{align*}
    EI_{i} &=\sum_{j=1}^{k_i} \frac{1}{\langle k \rangle} \cdot \log_{2}\Bigg(\dfrac{\dfrac{1}{\langle k \rangle}}{\dfrac{1}{N}}\Bigg) = \log_{2} \Big(\dfrac{N}{pN}\Big) \\
    EI_{ER} &=\dfrac{1}{N} \cdot  \sum_{i}^{N} -\log_{2}(p) = -\log_{2}(p) \tag{10}
\end{align*}

\subsubsection{\label{sec:SI_derivations_ring}Derivation: effective information of ring-lattice and star networks}

Here, we compare two classes of networks with the same average degree---ring lattice networks and star, or hub-and-spoke, graphs (see Fig. \ref{fig:star_ring}). In each network, we assume an average degree $\langle k \rangle = 2d$, with $d$ being the dimension. The $EI$ of star network, $EI_{star}$, approaches $0.0$ as $N$ gets larger, while the $EI$ of ring lattices approaches $\log_2(N) - \log_2(2d)$. These derivations are shown below, first for the \textit{d}-dimensional ring lattice, $EI_{d}$.

As every node in a ring lattice is connected to its $2d$ neighbors, each element of $\langle W^{out}_{i}\rangle$ is $\frac{1}{2d}$ and each element of $W^{in}$ is $\frac{2d}{2d\times N} = \frac{1}{N}$.
\begin{align}
    EI_{d} &= \dfrac{1}{N} \cdot  \sum_{i} \text{D}_{_{KL}}\Big[\big(\dfrac{1}{2d}, \dfrac{1}{2d}, ...\big) || \big(\dfrac{1}{N}, \dfrac{1}{N}, ...\big) \Big] \tag{11}
\end{align}

Each node in a \textit{d}-dimensional ring lattice is expected to be identical, so calculating the expected \textit{effect information}, $EI_i$, is equivalent to calculating the network's $EI$. As such, we observe:
\begin{align}
    EI_{i} &=  \sum_{j=1}^{2d} \frac{1}{2d} \cdot \log_{2}\Bigg(\dfrac{\dfrac{1}{2d}}{\dfrac{1}{N}}\Bigg) =  \log_{2}\Bigg(\dfrac{N}{2d}\Bigg) \nonumber \\
    EI_{d} &= \log_{2}(N) - \log_{2}(2d) \tag{12}
\end{align}

Note: the $EI$ of ring lattice networks reduces to simply the \textit{determinism} of the network. The $EI$ of ring lattice networks scale logarithmically with the size of the network, which is contrasted by the behavior of $EI$ in star networks. Star networks have a hub-and-spoke structure, where $N-1$ nodes of degree $k_{spoke}=1$ are connected a hub node, which itself has degree $k_{hub}=N-1$. For star networks, $EI$ approaches $0.0$ as the number of nodes increases. This derivation is shown below.
\begin{align*}
    EI_{star} = \dfrac{1}{N}\cdot\Bigg[ \sum_{i=1}^{N-1} &\text{D}_{_{KL}}\Big[W^{out}_{spoke} || \langle W^{out}_{i}\rangle \Big] +\nonumber\\
    &\text{D}_{_{KL}}\Big[W^{out}_{hub} || \langle W^{out}_{i}\rangle \Big]\Bigg]
\end{align*}

Every spoke has an out-weight vector $W^{out}_{i}$ with $N-1$ elements of $w_{ij}=0.0$ and one with $w_{ij} = 1.0$. The single hub, however, has $N-1$ elements of $w_{ij}=\frac{1}{N-1}$ with a single $w_{ij}=0.0$. Similarly, $\langle W^{out}_{i}\rangle$ consists of $N-1$ elements with values $\frac{1}{N(N-1)}$.
\begin{align}
    EI_{star} = \dfrac{1}{N}\cdot\Bigg[ \sum_{i=1}^{N-1} &\text{D}_{_{KL}}\Big[\dfrac{1}{1} || \dfrac{N-1}{N}\Big] +\nonumber\\ &\text{D}_{_{KL}}\Big[\dfrac{1}{N-1} || \dfrac{1}{N(N-1)}\Big]\Bigg] \tag{13}
\end{align}

\begin{figure*}[t]
    \centering
    \includegraphics[width=0.9\textwidth]{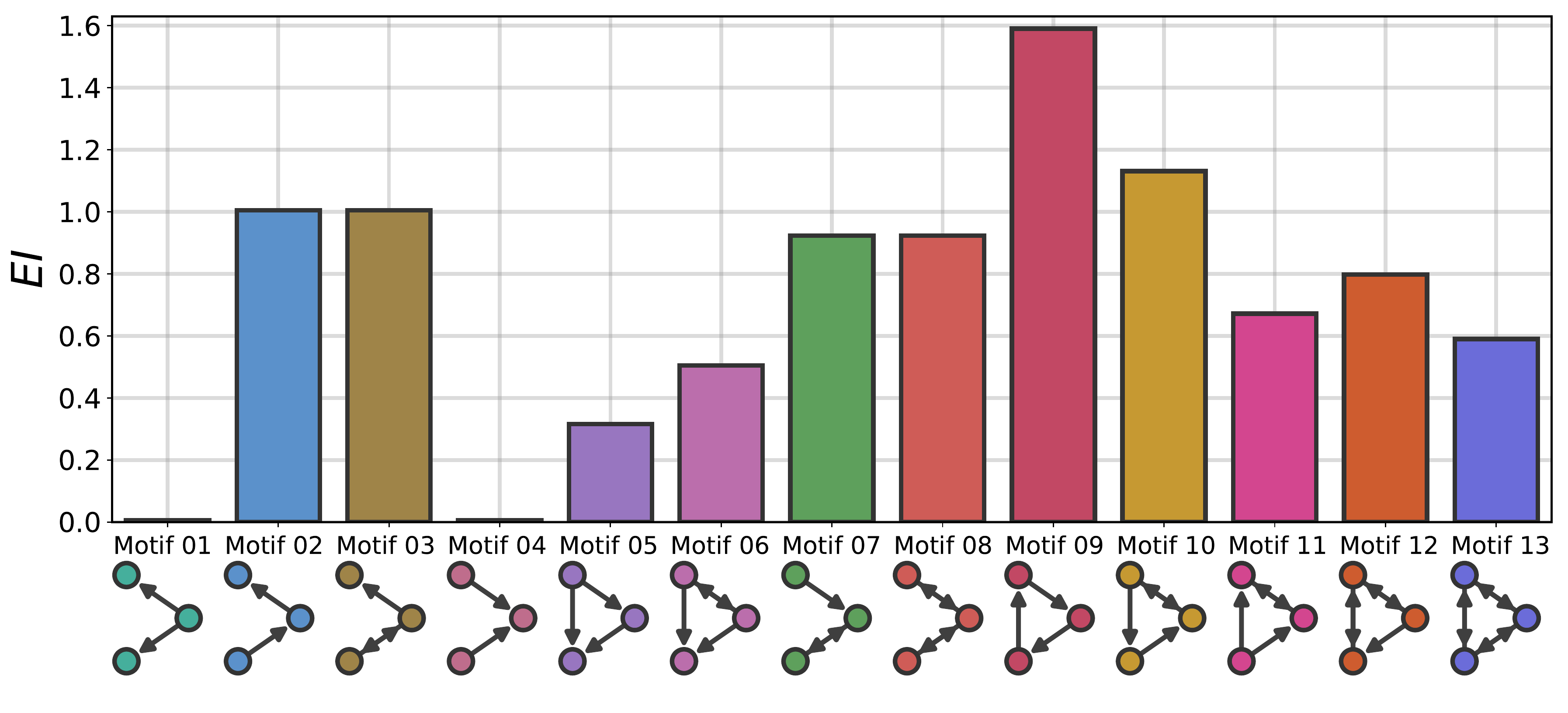}
    \caption{\textbf{Effective Information of network motifs.} All directed 3-node subgraphs and their $EI$.}
    \label{fig:networkmotifs}
\end{figure*}

Using the same techniques as above, this equation reduces to:
\begin{align}
    EI_{star} &= \dfrac{1}{N}\cdot\Bigg[(N-1)\cdot\log_{2}\Bigg(\dfrac{\dfrac{1}{1}}{\dfrac{(N-1)}{N}}\Bigg) +\nonumber\\ &\hspace{2.75cm}\log_{2}\Bigg(\dfrac{\dfrac{1}{N-1}}{\dfrac{1}{N(N-1)}}\Bigg)\Bigg] \nonumber \\
    EI_{star} &= \dfrac{N-1}{N}\cdot\log_{2}\Big(\frac{N}{N-1}\Big) + \dfrac{1}{N}\cdot\log_{2}\big(N\big) \nonumber \\ 
    EI_{star} &= 0.0 \hspace{0.2cm} \text{as} \hspace{0.15cm} \lim_{N\to\infty} \tag{14}
\end{align}

\begin{figure*}[t!]
    \centering
    \includegraphics[width=0.925\textwidth]{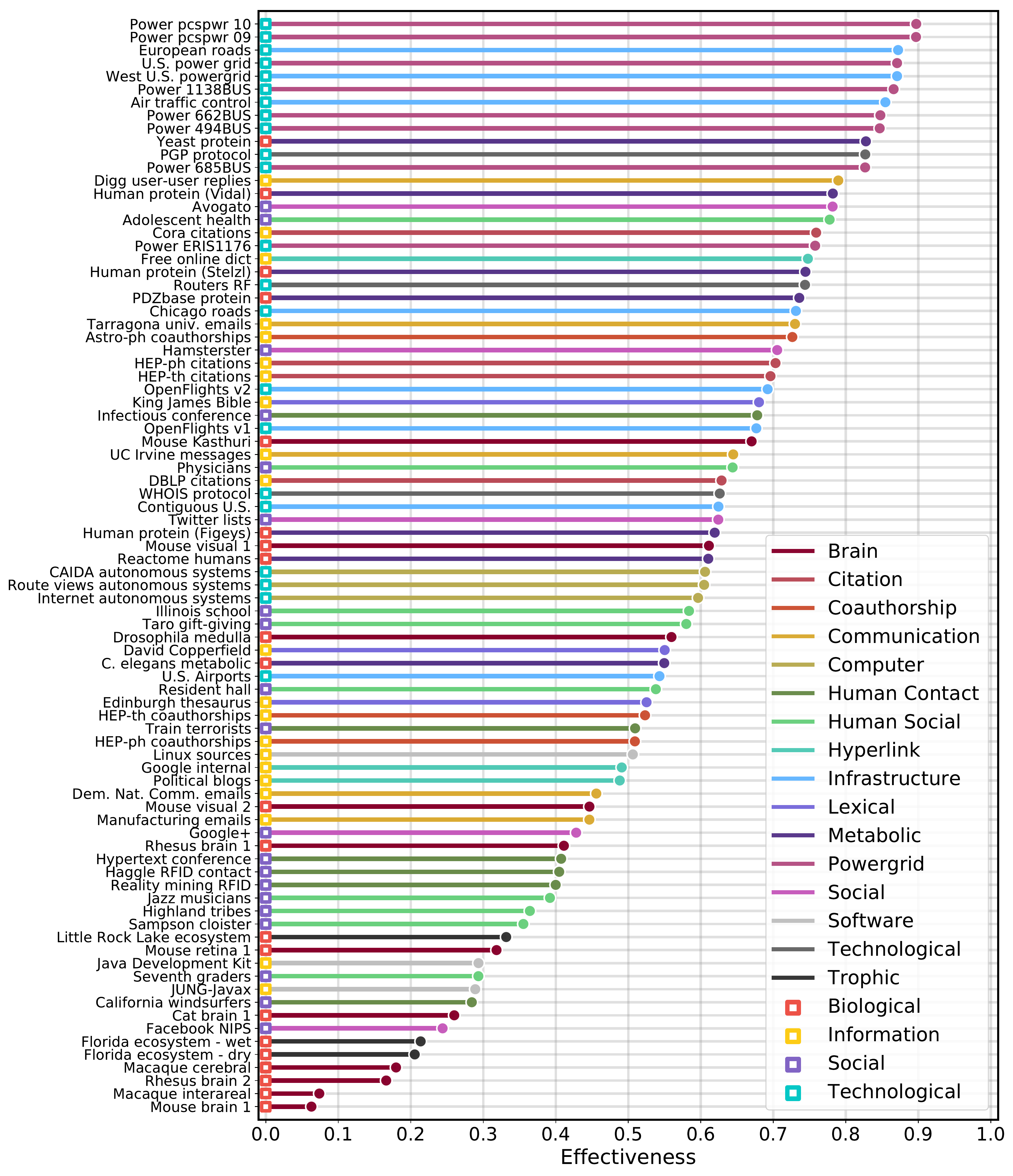}
    \caption{\textbf{Effectiveness of real networks.} Full data behind the results summarized in Fig. \ref{fig:ei_real_nets}, color-coded in two ways. First by 16 ``Domains'' (as in Table \ref{table_RealNetworkData}), which corresponds to the classification of each network from its source repository (in this case, the Konect database \protect\cite{Kunegis2013} or the Network Repository \protect\cite{Rossi2015}). The second categorization we report---those used in Fig. \ref{fig:ei_real_nets}---involves grouping the Domains into four ``Categories'' (``Cat.'' in Table \ref{table_RealNetworkData}): Biological, Information, Social, and Technological. These correspond to the colored squares to the right of each network's name.}
    \label{fig:Konect_full}
\end{figure*}

\begin{figure*}[t]
    \centering
    \includegraphics[width=0.99\textwidth]{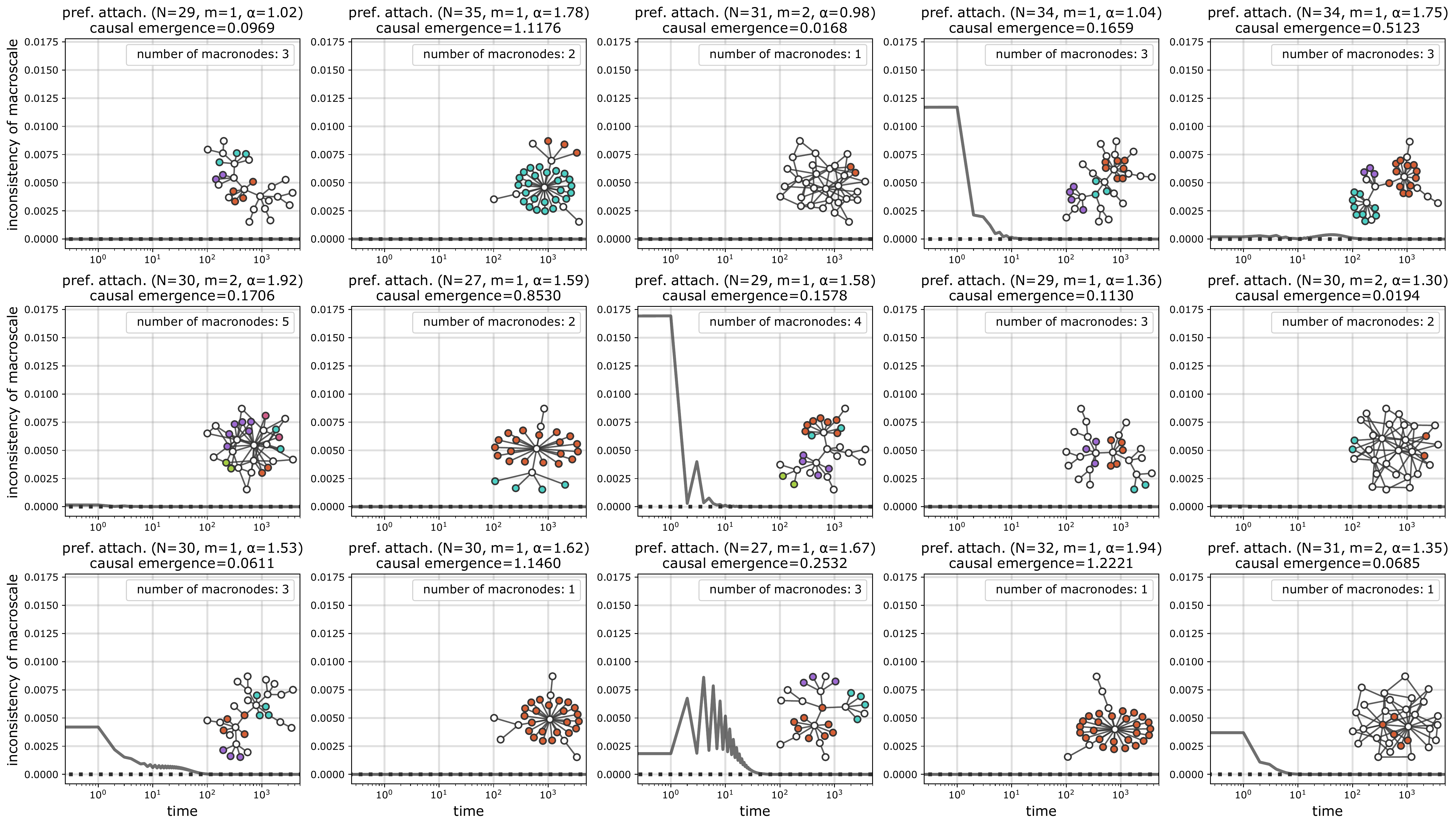}
    \caption{\textbf{Typically minimal inconsistency of higher-order macro-nodes.}  Each inset is of the microscale network, where each node's color corresponds to the $\mu | \pi$ macro-node it has been mapped to following one instance of the greedy algorithm detailed in the Materials \& Methods section. White nodes indicate a micro-node that was not grouped into a new macro-node. Inconsistency is plotted over time.}
    \label{fig:15inaccuracies}
\end{figure*}

\subsection{\label{sec:motifs}Network motifs as causal relationships} 

It is important to understand why certain motifs have more $EI$ while others have less. In Fig. \ref{fig:networkmotifs}, we show the $EI$ in 13 directed three-node network motifs. The connectivity of each motif drastically influences the $EI$. Motif 09---the directed cycle---is the motif with the highest $EI$. Intuitively, this fits with our definition of $EI$: the amount of certainty in the network (notably, each link in Motif 09, if taken to represent a causal relationship, is both necessary and sufficient). A random walker in this system has zero entropy (even if the direction of its path were reversed), whereas every other three-node motif does not contain that degree of certainty. Second, we see that Motif 04---a system with a ``sink'' node---has no $EI$, suggesting that a causal structure with that architecture is not informative, since all causes lead to the same effect. Similarly, because there are no outputs from two nodes in Motif 01, we see an $EI$ value of zero.

\subsection{\label{sec:SI_real_network_table}Table of network data}

In Table \ref{table_RealNetworkData}, we report the name, domain, source, category, and description of each of the 84 networks used in our comparison of $EI$ in real networks. These networks were selected primarily from the Konect database \protect\cite{Kunegis2013}, with supplemental datasets added from NetworkRepository \protect\cite{Rossi2015} when the Konect database lacked a sufficient number of datasets in a given category, since the two databases already significantly overlapped. In many cases, the interactions among nodes in these networks (i.e., their edges) can reasonably be interpreted as causal, directed influence, or dependencies such that the behavior of a node, $v_i$, at a given time can be thought to impact the behavior of its neighbors, $v_j$. By instituting relatively minimal requirements for selecting the above networks, we are able to assess the $EI$ in a variety of complex systems across different domains. However, while we can measure the $EI$ of any given network, the further interpretation of this $EI$ depends also on what the nodes and edges of a network represent. In a case where the nodes represent states of a system, such as a Markov process, then the $EI$ directly captures the information in the causal structure. In the case where the nodes represent merely dependencies or influence, $EI$ can still be informative as a metric to compare different networks. In a network specifically composed of non-causal correlations, then $EI$ is merely a structural property of the network's connectivity.

\begin{table*}[t!]
    \centering
    \begin{tabular}{| l | l | l | l | l |}
        \hline
        \textbf{Network name} & \textbf{Domain} & \textbf{Source} & \textbf{Cat.} & \textbf{Description} \\
        \hline \hline
        HEP-th citations & Citation & Konect & Inf. & high-energy physics (HEP) citations - theory \\ \hline
        HEP-ph citations & Citation & Konect & Inf. & HEP citations - phenomenology \\ \hline
        Cora citations & Citation & Konect & Inf. & citations from the Cora database \\ \hline
        DBLP citations & Citation & Konect & Inf. & database of scientific publications \\ \hline
        Astro-ph coauthorships & Coauthorship & Konect & Inf. & coauthors on astronomy arXiv papers \\ \hline
        HEP-th coauthorships & Coauthorship & Konect & Inf. & coauthors on HEP-theory arXiv papers \\ \hline
        HEP-ph coauthorships & Coauthorship & Konect & Inf. & coauthors on HEP-phenomenology arXiv papers \\ \hline
        Tarragona univ. emails & Communication & Konect & Soc. & emails from the University Rovira i Virgili \\ \hline
        Dem. Nat. Comm. emails & Communication & Konect & Soc. & 2016 Democratic National Committee email leak \\ \hline
        Digg user-user replies & Communication & Konect & Soc. & reply network from the social news website Digg \\ \hline
        UC Irvine messages & Communication & Konect & Soc. & messages between students at UC Irvine \\ \hline
        Manufacturing emails & Communication & Konect & Soc. & internal emails between employees at a company \\ \hline
        CAIDA autonomous systems & Computer & Konect & Tec. & autonomous systems network from CAIDA, 2007 \\ \hline
        Route views autonomous systems & Computer & Konect & Tec. & autonomous systems network \\ \hline
        Internet autonomous systems & Computer & Konect & Tec. & connected IP routing \\ \hline
        Haggle RFID contact & Human Contact & Konect & Soc. & human proximity, via carried wireless devices \\ \hline
        Reality mining RFID & Human Contact & Konect & Soc. & RFID data from 100 MIT students' interactions \\ \hline
        California windsurfers & Human Contact & Konect & Soc. & contacts between windsurfers California, 1986 \\ \hline
        Train terrorists & Human Contact & Konect & Soc. & contacts between Madrid train bombing suspects \\ \hline
        Hypertext conference & Human Contact & Konect & Soc. &  face-to-face contacts at the ACM Hypertext 2009 \\ \hline
        Infectious conference & Human Contact & Konect & Soc. & face-to-face contacts at INFECTIOUS, 2009 \\ \hline
        Jazz musicians & Human Social & Konect & Soc. & collaboration network between Jazz musicians \\ \hline
        Adolescent health & Human Social & Konect & Soc. & surveyed students list their best friends \\ \hline
        Physicians & Human Social & Konect & Soc. & innovation spread network among 246 physicians \\ \hline
        Resident hall & Human Social & Konect & Soc. & friendship ratings between students in a dorm \\ \hline
        Sampson cloister & Human Social & Konect & Soc. & relations between monks in a monastery \\ \hline
        Seventh graders & Human Social & Konect & Soc. & proximity ratings between seventh grade students \\ \hline
        Taro gift-giving & Human Social & Konect & Soc. & gift-givings (taro) between households \\ \hline
        Dutch college & Human Social & Konect & Soc. & friendship ratings between university freshmen \\ \hline
        Highland tribes & Human Social & Konect & Soc. & tribes in the Gahuku-Gama alliance structure \\ \hline
        Illinois school & Human Social & Konect & Soc. & friendships between boys at an Illinois highschool \\ \hline
        Free online dict. & Hyperlink & Konect & Inf. & cross references in Free Online Dict. of Computing \\ \hline
        Political blogs & Hyperlink & Konect & Inf. & hyperlinks between blogs, 2004 US election \\ \hline
        Google internal & Hyperlink & Konect & Inf. & hyperlink network from pages within Google.com \\ \hline
        Air traffic control & Infrastructure & Konect & Tec. & USA's FAA, Preferred Routes Database \\ \hline
        OpenFlights v1 & Infrastructure & Konect & Tec. & flight network between airports, OpenFlights.org \\ \hline
        OpenFlights v2 & Infrastructure & Konect & Tec. & flight network between airports, OpenFlights.org \\ \hline
        Contiguous U.S. & Infrastructure & Konect & Tec. & 48 contiguous states and D.C. of the U.S. \\ \hline
        European roads & Infrastructure & Konect & Tec. & international E-road network, mainly in Europe \\ \hline
        Chicago roads & Infrastructure & Konect & Tec. & road transportation network of the Chicago region \\ \hline
        West U.S. powergrid & Infrastructure & Konect & Tec. & power grid of the Western U.S. \\ \hline
        U.S. Airports & Infrastructure & Konect & Tec. & flights between US airports in 2010 \\ \hline
        David Copperfield & Lexical & Konect & Inf. & network of common noun and adjective adjacencies \\ \hline
        Edinburgh thesaurus & Lexical & Konect & Inf. & word association network, collected experimentally \\ \hline
        King James Bible & Lexical & Konect & Inf. & co-occurrence between nouns in the Bible \\ \hline
    \end{tabular}
    \caption{\textbf{Network datasets.} Continued on the following page.}
    \label{table_RealNetworkData}
\end{table*}

\setcounter{table}{1}
\begin{table*}[t!]
    \centering
    \begin{tabular}{| l | l | l | l | l |}
        \hline
        \textbf{Network name} & \textbf{Domain} & \textbf{Source} & \textbf{Cat.} & \textbf{Description} \\
        \hline \hline
        C. elegans metabolic & Metabolic & Konect & Bio. & metabolic network of the \textit{C. elegans} roundworm \\ \hline
        Human protein (Figeys) & Metabolic & Konect & Bio. & interactions network of proteins in Humans \\ \hline
        PDZbase protein & Metabolic & Konect & Bio. & protein-protein interactions from PDZBase \\ \hline
        Human protein (Stelzl) & Metabolic & Konect & Bio. & interactions network of proteins in Humans \\ \hline
        Human protein (Vidal) & Metabolic & Konect & Bio. & proteome-scale map of Human protein interactions \\ \hline
        Yeast protein & Metabolic & Konect & Bio. & protein interactions contained in yeast \\ \hline
        Reactome humans & Metabolic & Konect & Bio. & protein interactions, from the Reactome project \\ \hline
        Avogato & Social & Konect & Soc. & trust network for users of Advogato \\ \hline
        Google+ & Social & Konect & Soc. & Google+ user-user connections \\ \hline
        Hamsterster & Social & Konect & Soc. & friendships between users of hamsterster.com \\ \hline
        Twitter lists & Social & Konect & Soc. & Twitter user-user following network \\ \hline
        Facebook NIPS & Social & Konect & Soc. & Facebook user-user friendship network \\ \hline
        Linux dependency & Software & Konect & Inf. & Linux source code dependency network \\ \hline
        J.D.K. dependency & Software & Konect & Inf. & software class dependencies, JDK 1.6.0.7 \\ \hline
        JUNG/javax dependency & Software & Konect & Inf. & software class dependencies, JUNG 2.0.1 \& javax \\ \hline
        Florida ecosystem - dry & Trophic & Konect & Bio. & food web in the Florida wetlands (dry season) \\ \hline
        Florida ecosystem - wet & Trophic & Konect & Bio. & food web in the Florida wetlands (wet season) \\ \hline
        Little Rock Lake ecosystem & Trophic & Konect & Bio. & food web of Little Rock Lake, Wisconsin \\ \hline
        WHOIS protocol & Technological & NetworkRepository & Tec. & dataset of internet routing registries \\ \hline
        PGP protocol & Technological & NetworkRepository & Tec. & trust protocol of private keys of internet users \\ \hline
        Routers RF & Technological & NetworkRepository & Tec. & traceroute network between routers via Rocketfuel  \\ \hline
        Cat brain 1 & Brain & NetworkRepository & Bio. & fiber tracts between brain regions of a cat \\ \hline
        Drosophila medulla & Brain & NetworkRepository & Bio. & neuronal network from the medulla of a fly \\ \hline
        Rhesus brain 1 & Brain & NetworkRepository & Bio. & collation of tract tracing studies in primates \\ \hline
        Rhesus brain 2 & Brain & NetworkRepository & Bio. & inter-areal cortical networks from a primate \\ \hline
        Macaque cerebral & Brain & NetworkRepository & Bio. & connections between cerebral cortex of a primate \\ \hline
        Macaque interareal & Brain & NetworkRepository & Bio. & inter-areal cortical networks from  a primate \\ \hline
        Mouse Kasthuri & Brain & NetworkRepository & Bio. & neuronal network of a mouse \\ \hline
        Mouse brain 1 & Brain & NetworkRepository & Bio. & calcium imaging of neuronal networks in a mouse  \\ \hline
        Mouse retina 1 & Brain & NetworkRepository & Bio. & electron microscopy of neurons in mouse retina \\ \hline
        Mouse visual 1 & Brain & NetworkRepository & Bio. & electron microscopy of visual cortex of a mouse \\ \hline
        Mouse visual 2 & Brain & NetworkRepository & Bio. & electron microscopy of visual cortex of a mouse \\ \hline
        Power 1138BUS & Powergrid & NetworkRepository & Tec. & power system admittance, via Harwell-Boeing \\ \hline
        Power 494BUS & Powergrid & NetworkRepository & Tec. & power system admittance, via Harwell-Boeing \\ \hline
        Power 662BUS & Powergrid & NetworkRepository & Tec. & power system admittance, via Harwell-Boeing \\ \hline
        Power 685BUS & Powergrid & NetworkRepository & Tec. & power system admittance, via Harwell-Boeing \\ \hline
        U.S. power grid & Powergrid & NetworkRepository & Tec. & electricity / power transmission network in the U.S.\\ \hline
        Power pcspwr 09 & Powergrid & NetworkRepository & Tec. & BCSPWR09 powergrid data via Harwell-Boeing \\ \hline
        Power pcspwr 10 & Powergrid & NetworkRepository & Tec. &  BCSPWR10 powergrid data via Harwell-Boeing \\ \hline
        Power ERIS1176 & Powergrid & NetworkRepository & Tec. & powergrid data via Erisman, 1973 \\ \hline
    \end{tabular}
    \caption{\textbf{Network datasets (continued).}}
\end{table*}

\subsection{\label{sec:SI_inacc_examples}Examples of consistent macro-nodes}

In Fig. \ref{fig:15inaccuracies} we display 15 different parameterizations of small networks grown under degree-based preferential attachment. Each plot shows to the inconsistency of the mapping from the microscale to the macroscale, in bits, which corresponds to the KL divergence of the distribution of random walkers on microscale nodes and the same distribution at the macroscale. Each of these networks are consistent after 1000 timesteps, with eight showing full consistency from the start. These 15 example networks also show the range of causal emergence values that is found in networks.
 
\begin{figure*}[t!]
    \centering
    \includegraphics[width=0.75\textwidth]{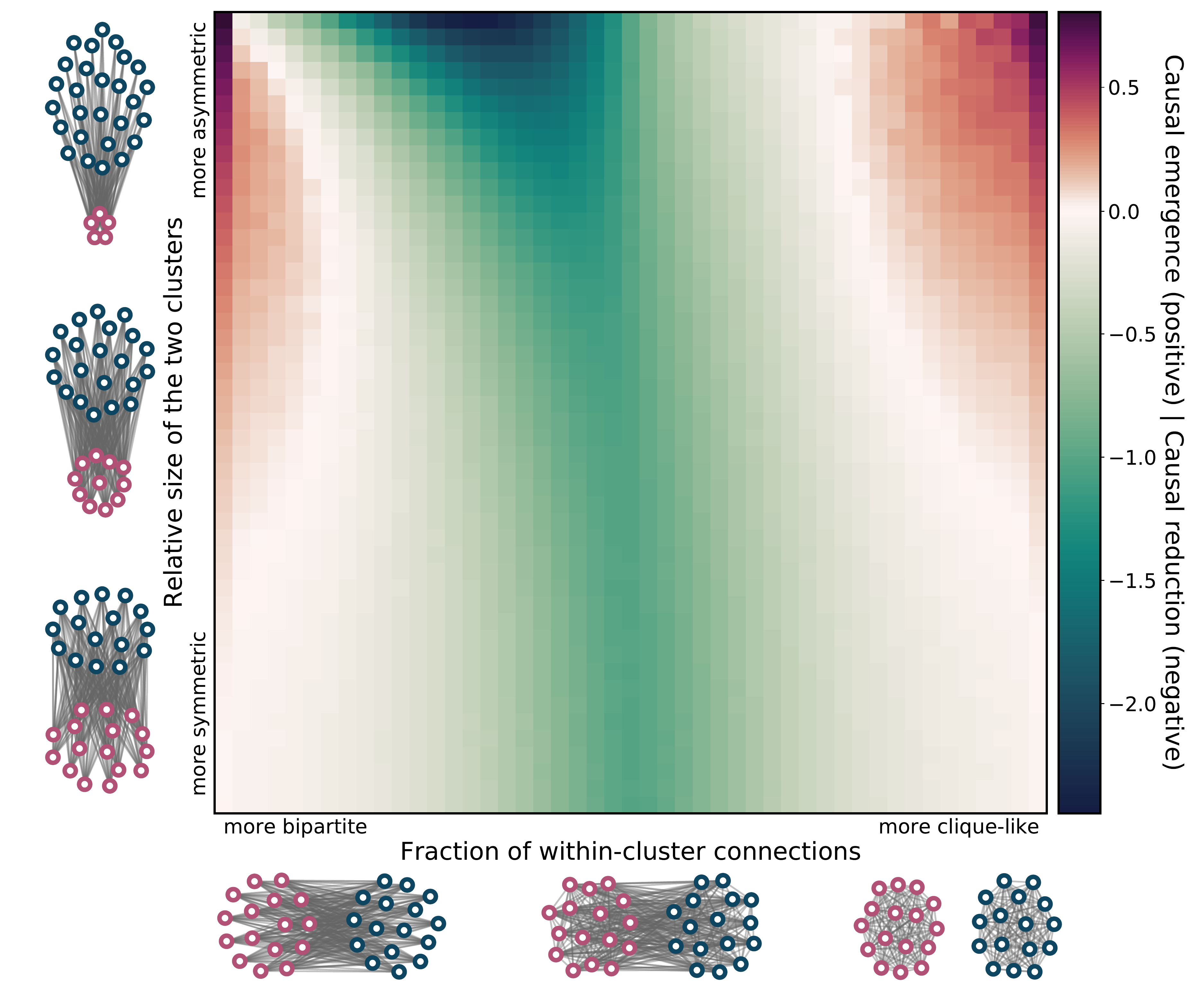}
    \caption{\textbf{Causal emergence in a simplified stochastic block model.} Schematic showing the role of the two relevant parameters---the fraction of nodes in each community (ranging from $r=0.50$ to $r<1.0$) and the fraction of within-cluster connections (ranging from $p=0.0$, a fully bipartite network, to $p=1.0$---two disconnected cliques). By repeatedly simulating networks under various combinations of parameters ($N=100$ with 100 simulations per combination of parameters), we see combinations that are more apt to produce networks with causal emergence.}
    \label{fig:sbm_ce}
\end{figure*}

\begin{figure*}[t!]
    \centering
    \subfloat[\label{fig:CE_ER_p}]{
        \includegraphics[width=0.9\columnwidth]{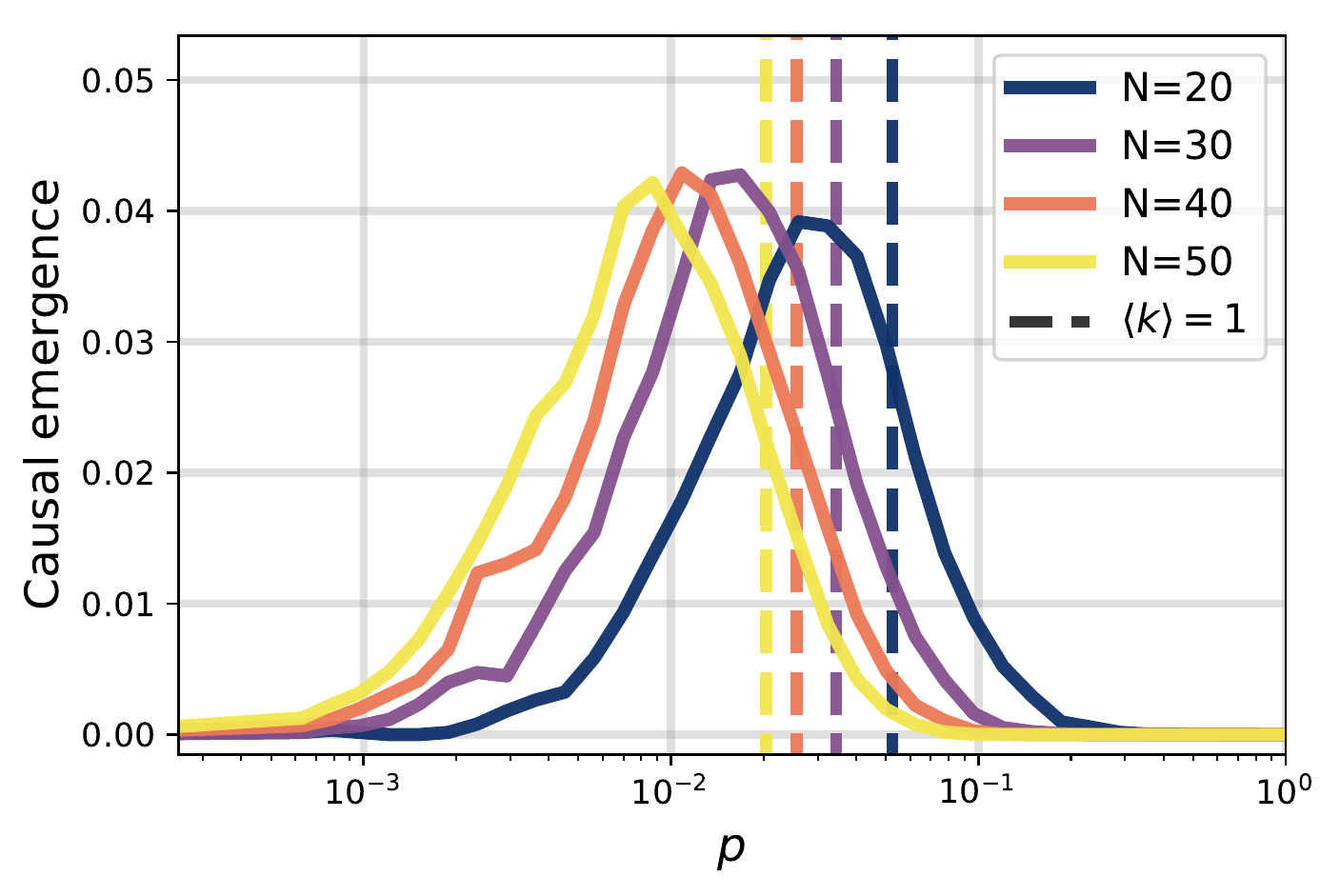}
    }\hfill
    \subfloat[\label{fig:CE_ER_k}]{
        \includegraphics[width=0.9\columnwidth]{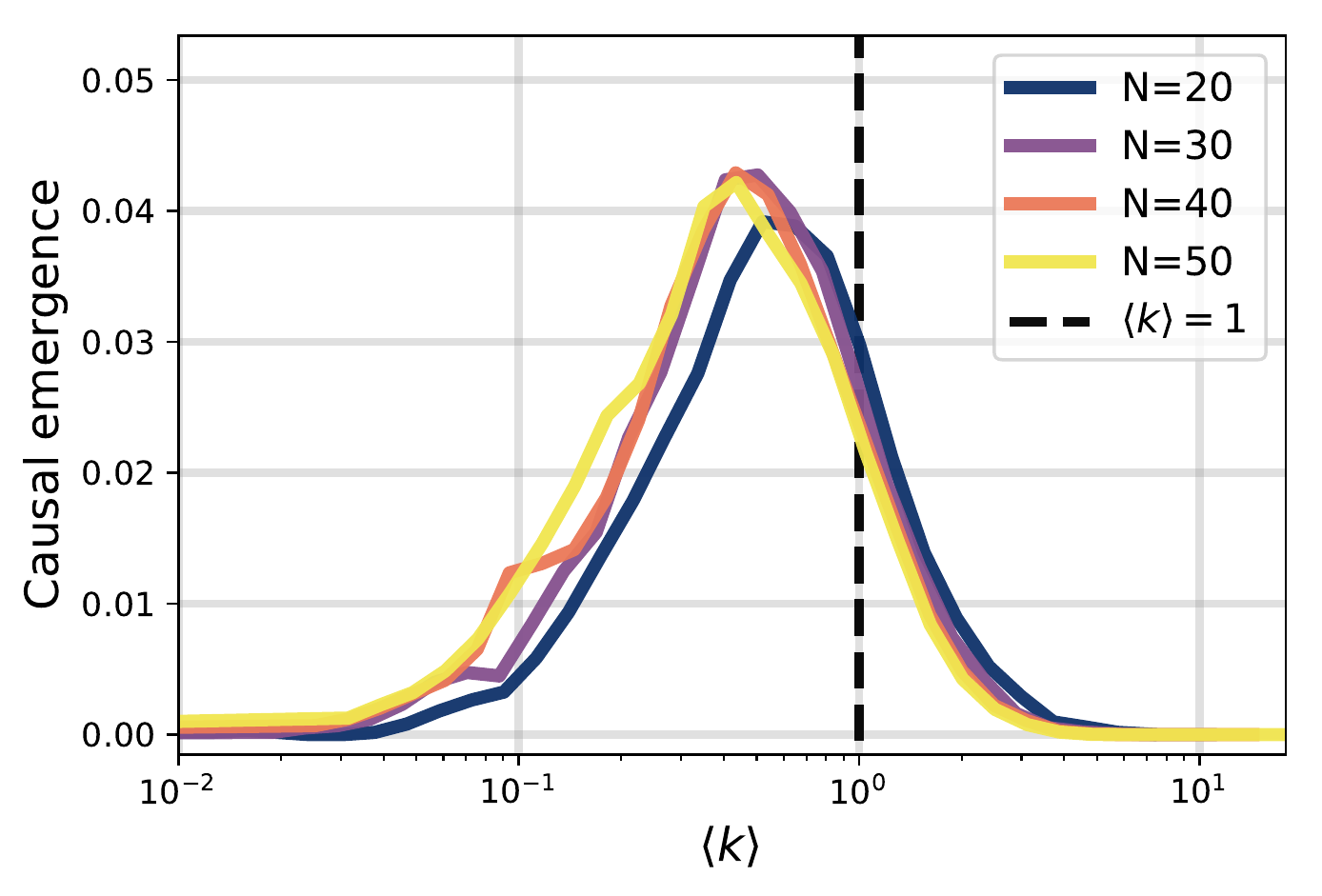}
    }
    \caption{\textbf{Causal emergence in Erd\H{o}s-R\'enyi networks.} (\textbf{A}) As the edge density, $p$, of ER networks increases and $N$ is held constant, the amount of causal emergence quickly drops to zero. (\textbf{B}) This drop occurs well before $pN = \langle k \rangle=1$, meaning the algorithm for uncovering causal emergence is only grouping small, disconnected, tree-like subgraphs that have yet to form into a giant component. Of note here is the low magnitude of causal emergence even in cases where the random network is not a single large component, and the vanishing of causal emergence after it is.}
    \label{fig:CE_ER}
\end{figure*}

\subsection{\label{sec:SI_characterize_CE}Emergent subgraphs}

What sort of subgraph connectivity leads to causal emergence? To explore this we take two independent subgraphs, and couple them together while varying their size, moving from clique-like to bipartite connectivity. We then check to see if grouping those clusters into macro-nodes leads to causal emergence (Fig. \ref{fig:sbm_ce}). Specifically, we simulate many small unweighted, undirected networks ($N=100$) from a stochastic block model with two clusters, and we vary the probability of within-cluster edges (from 0.0 to 1.0) as well as the size-asymmetry of the two clusters (illustrated around the border of Fig. \ref{fig:sbm_ce}). In each simulation, we group the microscale network into two macro-nodes, each corresponding to one cluster. What we observe is a causal emergence landscape with several important characteristics (Fig. \ref{fig:sbm_ce}). First, in these networks we observe causal emergence when the fraction of within-cluster connections is either very high or very low (right and left sides of the heatmap in Fig. \ref{fig:sbm_ce}). These are the conditions in which there is a large amount of uncertainty, or noise, in that subgraph. Not only that, however, causal emergence is most likely when there is a size asymmetry between the two clusters, suggesting that macroscales that maximize a network's $EI$ often do so by creating a more evenly distributed $\langle W^{out}_{i}\rangle$. In general, however, the space of subgraphs leans toward causal reduction (a loss of $EI$ after grouping), which fits with the success of reduction historically and explains why researchers and modelers should generally be biased toward reduction.

In cases of complete noise, with no asymmetries or differences between intra- or inter-connectivity between subgraphs, we should expect causal emergence to be impossible. Indeed, this is what we see for many parameterizations of Erd\H{o}s-R\'enyi networks of various sizes (Fig. \ref{fig:CE_ER}). This result follows from insights in Fig. \ref{fig:ei_p}, where the $EI$ of ER networks converges to a fixed value of $-\log_2(p)$ as the size of the network increases. Here, we observe some causal emergence in ER networks but only when the networks are very small. Importantly, the amount of causal emergence is also very small, especially relative to the causal emergence in networks with preferential attachment. This further suggests that causal emergence moves the existent structure of the network into focus by examining the network at a certain scale, rather than creating that structure from nothing.

\end{document}